\documentclass[11pt,a4paper]{article}
\usepackage[a4paper]{geometry}
\usepackage{amssymb,latexsym,amsmath,amsfonts,amsthm}
\usepackage{graphicx}
\usepackage{epstopdf}
\usepackage{tikz}
\usepackage{pgflibraryshapes}
\usetikzlibrary{arrows,decorations.markings}
\usepackage{subfigure}
\usepackage{overpic}
\usepackage{fullpage}
\usepackage{comment,verbatim}
\usepackage{hyperref}
\usepackage[title,titletoc]{appendix}
\usepackage{float}
\usepackage{appendix}

\newtheorem{thm}{Theorem}[section]
\newtheorem{lem}[thm]{Lemma}

\newtheorem{corollary}[thm]{Corollary}

\theoremstyle{remark}
\newtheorem{rem}[thm]{Remark}

\numberwithin{equation}{section}

\def\Im {\mathop{\rm Im}\nolimits}
\def\arg {\mathop{\rm arg}\nolimits}
\def\Re {\mathop{\rm Re}\nolimits}

\begin{document}

\title{ Singular asymptotics for the Clarkson-McLeod solutions of the fourth Painlev\'e equation}

\author{Jun Xia\footnotemark [1], ~Shuai-Xia Xu\footnotemark [2] ~and Yu-Qiu Zhao\footnotemark [1]}

\renewcommand{\thefootnote}{\fnsymbol{footnote}}
\footnotetext [1]  { Department of Mathematics, Sun Yat-sen University, GuangZhou 510275, China.}
\footnotetext [2] {Institut Franco-Chinois de l'Energie Nucl\'{e}aire, Sun Yat-sen University,
Guangzhou 510275, China.}

\date{}
\maketitle

\begin{abstract}
  We consider the Clarkson-McLeod solutions of the fourth Painlev\'e equation. This family of solutions behave like $\kappa D_{\alpha-\frac{1}{2}}^2(\sqrt{2}x)$ as $x\rightarrow +\infty$, where $\kappa $ is an arbitrary real constant and $D_{\alpha-\frac{1}{2}}(x)$ is the parabolic cylinder function.  Using the Deift-Zhou nonlinear steepest descent method,   we obtain
   the singular asymptotics of the solutions as $x\to-\infty$ when
   $\kappa \left( \kappa -\kappa ^*\right )>0$ for  some real constant  $\kappa ^*$.
    The   connection formulas are also explicitly evaluated. This proves  and extends Clarkson and McLeod's conjecture that when the parameter $\kappa >\kappa ^*>0$, the Clarkson-McLeod solutions have  infinitely many simple poles on the negative real axis. 
\end{abstract}

\noindent
\section{Introduction and statement of results}
We study the asymptotics of the solutions $q(x)$  of the fourth Painlev\'e (PIV, \cite{FIKN,NIST}) equation
\begin{equation}\label{PIV}
\frac{d^2 q}{ d x^2} =\frac{1}{2q}\left(\frac{dq}{dx}\right )^{2}+\frac{3}{2}q^{3}+4xq^{2}+\left(2x^{2}-4\alpha+\beta\right )q
-\frac{\beta^{2}}{2q},
\end{equation}
with  the parameters $\alpha \in \mathbb{R}$, $\beta=0$ and  satisfying the  boundary condition
\begin{equation}\label{Boundcondit}
q(x)\rightarrow 0\quad \mathrm{as}\quad x\rightarrow +\infty.
\end{equation}

In the pioneering works of Clarkson and McLeod \cite{CM}  and Bassom \emph{et al.} \cite{BCHM}, it is proven that
any real solution of \eqref{PIV} satisfying the boundary condition \eqref{Boundcondit} has the following asymptotic behavior
\begin{equation}\label{qAsy}
q(x;\kappa )\sim \kappa D^{2}_{\alpha-\frac{1}{2}}(\sqrt{2}x),~~~~x\to +\infty
\end{equation}
for some constant $\kappa $, where $D_{\nu}(x)$ is the parabolic cylinder function with order $\nu$; cf. \cite[Chapter 12]{NIST}. Conversely, for any real constant $\kappa $, there exists a unique solution of \eqref{PIV}  asymptotic to $\kappa D^{2}_{\alpha-\frac{1}{2}}(\sqrt{2}x)$ as $x\to+\infty$. %This family of
These solutions $q(x;\kappa )$ are now known as the \emph{Clarkson-McLeod solutions} of the fourth Painlev\'e  equation.  It is worth mentioning that a parameter $k$ is used in \cite{BCHM,CM} such that  $\kappa=2^{3/2} k^2$.

For  the  asymptotics of the  {Clarkson-McLeod solutions} as $x\to-\infty$,  there has been the following  conjecture.
\subsection*{Conjecture (Clarkson-McLeod \cite{CM})}\label{cojecture}
There exists a constant $\kappa ^*>0$ such that:
\begin{description}
\item{(a)} When $0<\kappa <\kappa ^{*}$, as $x\rightarrow-\infty$,
\begin{equation}\label{qAsy1}
q(x;\kappa )\sim c_n 2^{\alpha+1} x^{2 \alpha-1} e^{-x^{2}}%=c_n 2^{n+\frac{3}{2}} x^{2 n} e^{-x^{2}},
\end{equation}
if $\alpha-\frac{1}{2}=n \in \mathbb{N}$, and
\begin{equation}\label{qAsy2}
q(x;\kappa )\sim -\frac{2 x}{3}+(-1)^{\left[\alpha+\frac{1}{2}\right]} \frac{4 d}{\sqrt{3}} \sin \left(\frac{x^{2}}{\sqrt{3}}-\frac{4 d^{2}}{\sqrt{3}} \ln (-\sqrt{2} x)+a+O\left(\frac 1 {x^{2}}\right)\right)+O\left(\frac 1{x }\right)
\end{equation}
 if $\alpha-\frac{1}{2}\notin \mathbb{Z}$, where the constants $c_n$, $d$, $a$ are dependent on $\kappa $.
\item{(b)} When $\kappa =\kappa ^{*}$, $q(x;\kappa )$ behaves like $-2x$ as $x\rightarrow-\infty$.
\item{(c)} When $\kappa >\kappa ^{*}$, $q(x;\kappa )$ has a pole   on the negative real axis.
\end{description}

In case (a), the asymptotic formula \eqref{qAsy1} has been proven in \cite{BCHM,CM}  for $\alpha-\frac{1}{2} \in \mathbb{N}$, and the values of $c_n$ and $\kappa ^{*}$ were explicitly evaluated as
$$
c_n=\frac{\kappa }{2\sqrt{2}-2\sqrt{2\pi}\,n!\,\kappa },\qquad
\kappa ^*=\frac{1}{\sqrt{\pi}\, n!}.
$$
If $\alpha-\frac{1}{2}\notin \mathbb{Z}$, the value of $\kappa ^*$ was conjectured to be
\begin{equation}\label{Kstar}
\kappa ^*=\frac{1}{\sqrt{\pi}\, \Gamma\left(\alpha+\frac{1}{2}\right)}.
\end{equation}
The asymptotic formula \eqref{qAsy2} was later justified by Abdullayev \cite{Ab} using the integral equation method and by Its and Kapaev \cite{IK} via the isomonodromy method, respectively. The connection formulas for the dependence on $\kappa $ of $a$ and $d$ in  \eqref{qAsy2} were explicitly evaluated in \cite{IK,WZ}. However, to the best of our knowledge, the asymptotics of $q(x;\kappa )$ as $x\to -\infty$ in cases (b) and  (c) are still to be explored.

The present paper is devoted to the studies of the asymptotics  as $x\to -\infty$ of the  Clarkson-McLeod solutions
 corresponding to   case (c) of the Clarkson-McLeod conjecture.  We
derive the singular asymptotics  for this family of Clarkson-McLeod solutions as $x\rightarrow-\infty$  with   explicit expressions of the connection formulas.

\begin{thm}\label{thm}
Assume that $\alpha\in \mathbb{R}$, $\alpha-\frac{1}{2}\notin \mathbb{Z}$, $\beta=0$, $\kappa ^*$ be given by \eqref{Kstar}, and let $q(x;\kappa )$ be a real solution  of \eqref{PIV} satisfying the asymptotic behavior  \eqref{qAsy} as $x\rightarrow +\infty$  with real parameter $\kappa$ such that  $\kappa\left (\kappa -\kappa ^*\right )>0$,
then $q(x;\kappa )$ has the following asymptotic behavior as $x\rightarrow-\infty$
\begin{equation}\label{qasymp-infty}
q(x;\kappa )=-\frac{2}{3}x+\frac{2x}{2\cos\left(\frac{\sqrt{3}}{3}x^{2}-b\ln \left(2 \sqrt{3} x^{2}\right)+\psi\right)+1}+O\left(\frac{1}{x}\right),
\end{equation}
where
\begin{equation}\label{connectionformula}
\left\{
\begin{aligned}
b&=-\frac{1}{2\pi} \ln(|\rho|^{2}-1),\\
\psi&=-\frac{2\pi}{3}\alpha-\arg\Gamma\left(-bi+\frac{1}{2}\right)-\arg \rho,
\end{aligned}\right.
\end{equation}
and the connection between $\rho$ and $\kappa $ is given by
\begin{equation}\label{Kandsstar}
\rho=1-\frac{2\pi^{\frac{3}{2}}}{e^{\pi i\alpha}\Gamma\left(\frac{1}{2}-\alpha\right)}\kappa .
\end{equation}
The error term in the asymptotic expansion is uniform for $x$ bounded away from the singularities appearing on the right-hand side of \eqref{qasymp-infty}.
\end{thm}

\begin{rem}\label{remark1}
The existence of a family of solutions satisfying the  asymptotic behavior  \eqref{qAsy}  was first established by Bassom \emph{et al.} in \cite{BCHM} for the case $\kappa >0$ and subsequently proven by Its and Kapaev in \cite{IK} for general $\kappa \in \mathbb{R}$ with $\kappa \neq 0$. The remaining case $\kappa =0$ is corresponding to the trivial solution  $q(x; 0)=0$.
\end{rem}

From the asymptotic formula \eqref{qasymp-infty}, we obtain the following asymptotic approximation of the location of the  large negative poles of  the solution $q(x;\kappa )$.
\begin{corollary}\label{cor} Under the assumptions  of Theorem \ref{thm}, the solution  $q(x;\kappa )$
  has infinitely many simple poles on the negative real axis. Moreover, we have the following  asymptotic approximation of the locations of  large negative poles of  $q(x;\kappa )$
\begin{equation}\label{poles}
a_n^{\pm}=- (2\pi)^{\frac 1 2}  3^{\frac 1 4} \left [\sqrt n + \frac {b\ln n}{4\pi \sqrt n} +   \frac {b \ln (12\pi) -\psi\pm \frac {2\pi} 3} {4\pi\sqrt n}
+O\left (\frac{\ln^2 n}{ n^{3/2}}\right )
\right ], \quad n\to\infty,
\end{equation}
with $b$ and $\psi$ given in \eqref{connectionformula}.
\end{corollary}

Let us consider the case $\kappa^*>0$. For $\kappa >\kappa ^*$, the existence of a negative pole of the solution $q(x;\kappa )$ of the fourth Paivnlev\'e equation  \eqref{PIV} satisfying the asymptotic behavior  \eqref{qAsy}  was first conjectured by Clarkson and McLeod \cite{CM}.
For  $\kappa >\kappa ^*$ and  $\kappa <0$, it was shown numerically by  Reeger and Fornberg in \cite[Figures 5 and 7]{RF}  that  when $\alpha=0$, the solution $q(x;\kappa )$ have infinitely many  poles on the negative real axis. Corollary \ref{cor} rigorously
confirms and extends both the conjecture of Clarkson and McLeod, and the numerical results of Reeger and Fornberg, in that there are infinitely many poles of
$q(x;\kappa )$ on the negative real $x$-axis
for general parameter $\alpha$.
 Further  numerical analysis of \eqref{PIV} is worthwhile
to   demonstrate  the accuracy of the asymptotic  results.
Analogous to \eqref{qasymp-infty}, singular asymptotics of a family of solutions of homogeneous and inhomogeneous second Painlev\'e equation  have been
 established earlier in \cite{BI} and \cite{Hu}, respectively.

In Theorem \ref{thm},
the restriction $\alpha-\frac 1 2 \not\in  \mathbb{Z}$ has been brought in for technical reasons; cf. \eqref{E0}. However,
the  asymptotics of the Clarkson-McLeod solution $q(x;\kappa )$ as $x\rightarrow-\infty$  for
real $\kappa$ such that   $ \kappa \left (\kappa -\kappa ^*\right )\leq 0$,
along with the exceptional case $\alpha-\frac 1 2  \in  \mathbb{Z}$, can also be derived by using the Deift-Zhou nonlinear steepest descent method and we will report those results elsewhere.

The rest of the present paper is arranged as follows. In Section \ref{RHP}, we recall the  Riemann-Hilbert (RH) problem for the PIV equation \eqref{PIV}. The nonlinear steepest descent analysis of the RH problem are performed in Section
\ref{RHanalysis}.  The main results will then be proved in the final section, Section \ref{sec:proof}: Proof of Theorem \ref{thm} will be given in Section \ref{proof}, and proof of Corollary \ref{cor} in Section \ref{proof:cor}. For the convenience of the reader, we  collect in the Appendix the Airy, Bessel and parabolic cylinder parametrices used in the RH analysis.

\section{  Riemann-Hilbert problem for PIV equation}
\label{RHP}

We recall the  RH problem for the PIV equation \eqref{PIV} in this section. More details can be found in \cite[Section 2]{IK} and \cite[Chapter 5.1]{FIKN}.

\begin{figure}[t]
  \centering
  \includegraphics[width=7cm,height=7cm]{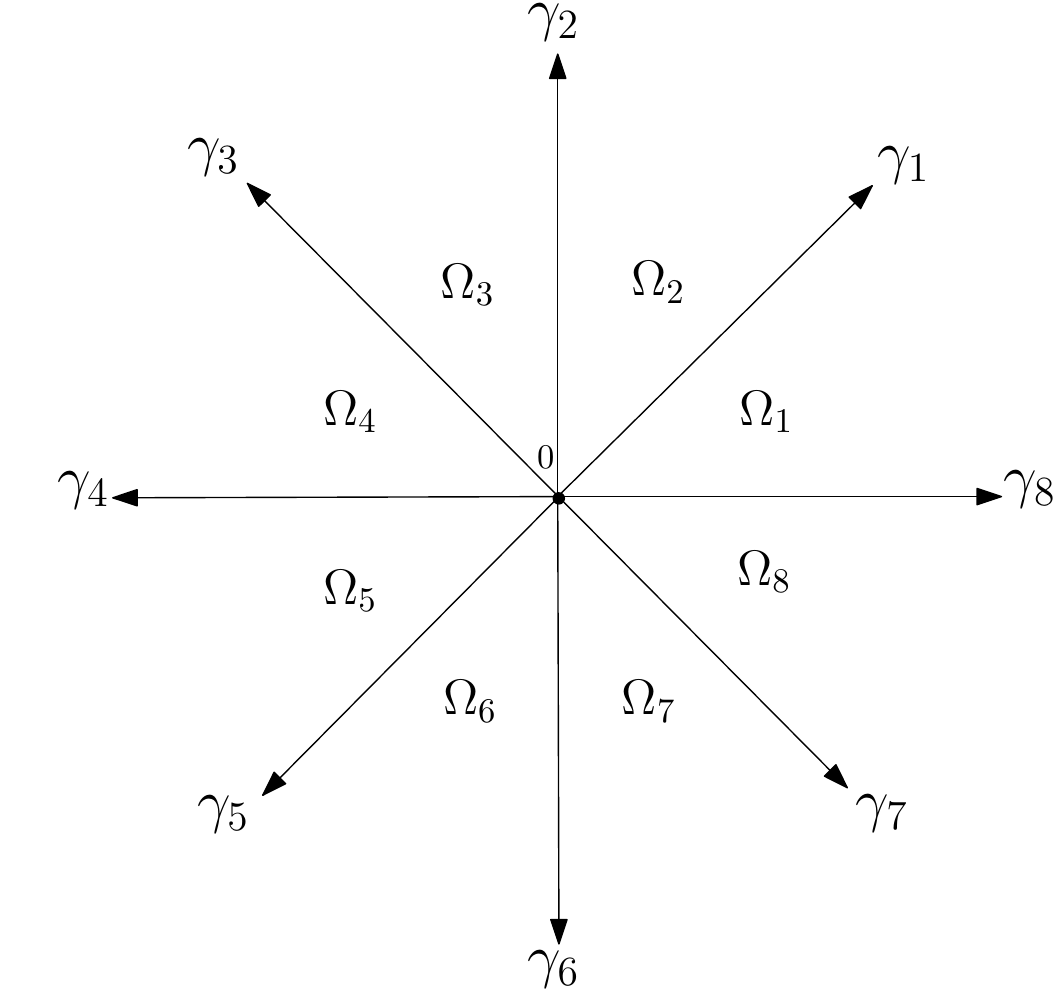}\\
  \caption{The jump curves $\Sigma$ of the RH problem for $\Psi(\xi)$}\label{PIVj}
\end{figure}

\subsection*{RH problem for PIV}
Let $\Sigma=\cup^{8}_{k=1}\gamma_{k}$, where $\gamma_{k}=\{\xi\in \mathbb{C} :  \arg\xi=k\pi/4\}$. Then, $\Psi(\xi):=\Psi(\xi;x)$ satisfies the following RH problem.
\begin{description}
\item{(1)} $\Psi(\xi)$ is analytic for $\xi\in\mathbb{C}\setminus \Sigma$, where $\Sigma$ is illustrated in Figure \ref{PIVj}.

\item{(2)} $\Psi(\xi)$ satisfies the   jump conditions
$$
\Psi_{+}(\xi)=\Psi_{-}(\xi)\left\{
\begin{aligned}
&S_{k},\quad &\xi&\in\gamma_{k},\ k=1,\cdots,7,\\
&S_{8}e^{-2\pi i(\alpha-\beta)\sigma_{3}},\quad &\xi&\in\gamma_{8},
\end{aligned}
\right.
$$
where $\Psi_{+}$ and $\Psi_{-}$ denote  the limits of the function $\Psi$ on the ray $\gamma_k$ from the left and the right hand side, respectively. Here $\sigma_3$ is one of the Pauli matrices
\begin{equation*}\label{Pauli}
\sigma_1=\begin{pmatrix}0 & 1\\ 1 & 0\end{pmatrix},\quad \sigma_2=\begin{pmatrix}0 & -i\\ i & 0\end{pmatrix},\quad \sigma_3=\begin{pmatrix}1 & 0\\ 0 & -1\end{pmatrix}.
\end{equation*}
The Stokes matrices  $S_k$'s  are of the form
\begin{equation*}
S_{2j-1}=
\begin{pmatrix}
1 & s_{2j-1}\\
0 & 1
\end{pmatrix},\quad
S_{2j}=
\begin{pmatrix}
1 & 0\\
s_{2j} & 1
\end{pmatrix}, \quad \ j=1,2,3,4.
\end{equation*}
The constants $s_k$'s are known as the Stokes multipliers. They are constrained by
\begin{equation}\label{srela1}
s_{k+4}=-s_{k}e^{(-1)^{k}2\pi i(\alpha-\beta)},\quad k=1,2,3,4,
\end{equation}
\begin{equation}\label{srela2}
\left [(1+s_{1}s_{2})(1+s_{3}s_{4})+s_{1}s_{4}\right ]e^{-\pi i(\alpha-\beta)}-(1+s_{2}s_{3})e^{\pi i(\alpha-\beta)}=-2i\sin\pi\alpha.
\end{equation}
\item{(3)} $\Psi(\xi)$ satisfies the following asymptotic behavior as $\xi\rightarrow\infty$
\begin{equation}\label{Asyatinfty}
\Psi(\xi)=\left(I+\frac{\Psi_{1}}{\xi}
+O\left(\frac{1}{\xi^{2}}\right)\right)
e^{\theta(\xi)\sigma_{3}},\quad \theta(\xi)=\frac{1}{8}\xi^{4}+\frac{1}{2}x\xi^{2}+(\alpha-\beta)\ln\xi,
\end{equation}
where the branch of $\ln\xi$ is chosen such that $\arg\xi\in(0,2\pi)$.
\item{(4)} $\Psi(\xi)$ has the following asymptotic behavior near $\xi=0$
\begin{equation}\label{PsiAsy0}
\Psi(\xi)=\Psi_0(\xi)\xi^{\alpha\sigma_{3}}E
\end{equation}for the generic case  $\alpha-\frac{1}{2}\notin \mathbb{Z}$,
where $\Psi_0(\xi)$ is analytic and invertible, the branch of $\xi^{\alpha}$ is chosen such that $\arg\xi\in(0,2\pi)$ and
\begin{equation}\label{E}
E=\left\{
\begin{aligned}
&E_0,\quad \det E_0=1,\quad &\xi&\in\Omega_1,\\
&E_0S_{1}\cdots S_{k-1},\quad &\xi&\in\Omega_{k},\quad k=2,\cdots,8.
\end{aligned}\right.
\end{equation}
\end{description}

The solution $q(x)$ of the PIV equation \eqref{PIV} is then determined by the solution to the above RH problem for $\Psi(\xi)$ via the   formula
\begin{equation}\label{solu1}
q(x)=(\Psi_{1})_{12}(\Psi_{1})_{21},
\end{equation}
where $\Psi_1=\Psi_1(x)$ is the coefficient in \eqref{Asyatinfty} and $M_{ij}$ denotes the ($i,j$)-th entry of a matrix $M$.

The above RH formulation is valid no matter $\beta=0$ or not. Nevertheless,
if $\beta\not=0$,
   the PIV equation \eqref{PIV} possesses no Clarkson-McLeod solutions. % in this case.
We focus on the asymptotic analysis of the Clarkson-McLeod solutions from now on, and   consider the case $\beta=0$.

From Its and Kapaev \cite[Equation (2.23)]{IK}, it is seen that for any real solution of \eqref{PIV}, the Stokes multipliers must satisfy the conditions
\begin{equation}\label{assum1}
\bar{s}_0=s_0,\quad \bar{s}_1=-s_{3}e^{2\pi i\alpha}.
\end{equation}
For any real solution satisfying asymptotic behavior \eqref{qAsy}, it follows from \cite[Equation (2.42)]{IK} that the associated Stokes multipliers further fulfill the following conditions
\begin{equation}\label{assum2}
s_2=0,\quad s_1+s_3=0,\quad s_*\neq1, \quad (1-s_*)e^{\pi i\alpha}\in\mathbb{R},
\end{equation}
where $s_*$ is defined by
\begin{equation}\label{sstar}
s_*=1+s_0s_1.
\end{equation}
The conditions $s_2=0$ and $s_1+s_3=0$ in \eqref{assum2}  imply that the Stokes matrices satisfy
$$S_2=S_6=I, \quad S_1=S_3^{-1}.$$
 Moreover, from \cite[Equation (3.15)]{IK}, the connection matrix $E_0$ takes the form
\begin{equation}\label{E0}
E_0=p^{\sigma_3}\begin{pmatrix}
1 & 0 \\ \frac{s_0e^{2\pi i\alpha}}{e^{2\pi i\alpha}+1} & 1
\end{pmatrix},
\end{equation}
where $p$ is an arbitrary nonzero constant.

In view of \cite[Equation (3.42)]{IK}, we have the following explicit relation between the parameter $\kappa $ in \eqref{qAsy} and the composite Stokes multiplier $s_*$
\begin{equation}\label{Krepre}
\kappa =\frac{e^{\pi i\alpha} \Gamma\left(\textstyle\frac{1}{2}-\alpha\right)}{2
\pi^{\frac{3}{2}}}(1-s_*).
\end{equation}
The relation \eqref{Krepre}, together with \eqref{Kstar}, \eqref{assum1} and \eqref{assum2},  implies that the conditions on $\kappa $  can be equivalently expressed in terms of the composite Stokes multiplier $s_*$ %defined in \eqref{sstar} as follows
as shown in the following table:
\begin{table}[h]
\centering
\begin{tabular}{|c|c|c|}
\hline
 $s_*$ & $\kappa^*>0$ & $\kappa^*<0$  \\
\hline
 $|s_*|<1$  & $0<\kappa <\kappa^*$    & $\kappa^*<\kappa<0$ \\
\hline
 $|s_*|=1,\ s_*\neq1$  & $\kappa=\kappa^*$  & $\kappa=\kappa^*$ \\
\hline
$|s_*|>1$ & $\kappa<0\ \mathrm{or} \ \kappa>\kappa^*$   & $\kappa<\kappa^*\ \mathrm{or} \ \kappa>0$  \\
\hline
\end{tabular}
\caption{~The correspondence
between $s_*$ and $\kappa$}\label{table}
\end{table}

\section{Nonlinear steepest descent analysis}
\label{RHanalysis}
In this section, we consider the case  %\eqref{case3} where
$ |s_*|>1$.  We shall  perform the Deift-Zhou nonlinear steepest descent analysis of the  RH  problem $\Psi$ for the PIV equation \eqref{PIV}  as $x\to-\infty$.

Assume now that $x<0$. We begin with the following re-scaling transformation
\begin{equation}\label{rescaling}
\Phi(z)=(-x)^{-\frac{\alpha}{2}\sigma_3}\Psi\left((-x)^{\frac{1}{2}}z;x\right ).
\end{equation}
As a result, $\Phi(z)$ satisfies the following RH problem.

\subsection*{RH problem for $\Phi(z)$}
\begin{description}
\item{(1)} $\Phi(z)$ is analytic for $z\in\mathbb{C}\setminus\left\{\Sigma\setminus(\gamma_2\cup\gamma_6)\right\}$; cf. Figure \ref{PIVj} for the contour.

\item{(2)} $\Phi(z)$ fulfills the following jump relations
$$
\Phi_{+}(z)=\Phi_{-}(z)\left\{
\begin{aligned}
&S_{k},\quad &z&\in\gamma_{k},\ k=1,3,4,5,7,\\
&S_{8}e^{-2\pi i\alpha\sigma_{3}},\quad &z&\in\gamma_{8}.
\end{aligned}
\right.
$$
\item{(3)} At infinity, $\Phi(z)$ has the following asymptotic behavior
\begin{equation}\label{Asyatinfty1}
\Phi(z)=\left(I+\frac{\Phi_{1}}{z}+O\left(\frac 1 {z^2}\right)\right)
z^{\alpha\sigma_3}
e^{ \frac{x^{2}}{8}\left(z^{4}-4z^{2}\right)\sigma_{3}},
\end{equation}
where the branch of $z^{\alpha}$ is chosen such that $\arg z\in(0,2\pi)$.
\item{(4)} $\Phi(z)$ has the same asymptotic behavior as $\Psi(z)$ at $z=0$; see \eqref{PsiAsy0} and \eqref{E}.
\end{description}
Simultaneously, it follows from \eqref{solu1} and \eqref{rescaling} that
\begin{equation}\label{solu2}
q(x;\kappa )=-x\left(\Phi_{1}\right)_{12}\left(\Phi_{1}\right)_{21},
\end{equation}
where $\Phi_1=\Phi_1(x)$ is the coefficient in the expansion \eqref{Asyatinfty1}.

\subsection{Normalization and deformations of the jump curves}
To normalize the asymptotic behavior of $\Phi(z)$ at infinity, we introduce the $g$-function
\begin{equation}\label{g-function}
g(z)=\frac{1}{8}z\left(z^{2}-\frac{8}{3}\right)^{\frac{3}{2}},
\end{equation}
where the branch of the power is taken such that $\arg(z\pm\textstyle\sqrt{\frac{8}{3}})\in(-\pi,\pi)$. A straightforward computation gives
\begin{equation*}
g(z)=\frac{1}{8}z^4-\frac{1}{2}z^2+\frac{1}{3}+O(z^{-2}), \quad  \mathrm{as}\quad z\rightarrow \infty.
\end{equation*}
It is easy to see that $g(z)$ has four saddle points, namely, the points satisfying $g'(z)=0$,
\begin{equation*}
z_{1,\pm}=\pm\sqrt{\frac{2}{3}},\quad z_{2,\pm}=\pm\sqrt{\frac{8}{3}}.
\end{equation*}

The second transformation is now defined as
\begin{equation}\label{thetahat}
U(z)=e^{\frac{x^2}{3}\sigma_{3}}\Phi(z)z^{-\alpha\sigma_3}e^{-x^2g(z)\sigma_{3}}.
\end{equation}
As a consequence, $U(z)$ solves the following RH problem.

\subsection*{RH problem for $U(z)$}
\begin{description}
\item{(1)} $U(z)$ is analytic for $z\in\mathbb{C}\setminus\left\{\Sigma\setminus(\gamma_2\cup\gamma_6)\right\}$; cf. Figure \ref{PIVj} for the contour.

\item{(2)} $U(z)$ satisfies $U_{+}(z)=U_{-}(z)J_U(z)$, where
$$J_U(z)=\left\{
\begin{aligned}
&\begin{pmatrix}1 & s_kz^{2\alpha}e^{2x^2g(z)}\\ 0 & 1 \end{pmatrix},\quad &z&\in\gamma_{k},\ k=1,3,5,7,\\
&\begin{pmatrix}e^{x^2(g_-(z)-g_+(z))} & 0\\ -s_0|z|^{-2\alpha}e^{-x^2(g_+(z)+g_-(z))} & e^{x^2(g_+(z)-g_-(z))} \end{pmatrix},\quad &z&\in\gamma_{4},\\
&\begin{pmatrix}e^{x^2(g_-(z)-g_+(z))} & 0\\ s_0|z|^{-2\alpha}e^{-x^2(g_+(z)+g_-(z))} & e^{x^2(g_+(z)-g_-(z))} \end{pmatrix},\quad &z&\in\gamma_{8}.
\end{aligned}
\right.
$$
\item{(3)} $U(z)$ is normalized at infinity, that is,
$$U(z)=I+O(z^{-1}),\quad \mathrm{as}\quad z\rightarrow\infty.$$
\item{(4)} $U(z)$ possesses the following asymptotic behavior as $z\to 0$
\begin{equation}\label{U0}
U(z)=U_0(z)z^{\alpha\sigma_{3}}E_0 z^{-\alpha\sigma_3}e^{-x^2g(z)\sigma_{3}},
\end{equation}
where  $\arg z\in(0, \frac{\pi}{4})$ and $U_0(z)$ is analytic in a neighborhood of $z=0$. The behavior of $U(z)$ in other regions  is determined by \eqref{U0} and the jump relations satisfied by $U(z)$.

\end{description}

\begin{figure}[H]
  \centering
  \includegraphics[width=11cm,height=6.875cm]{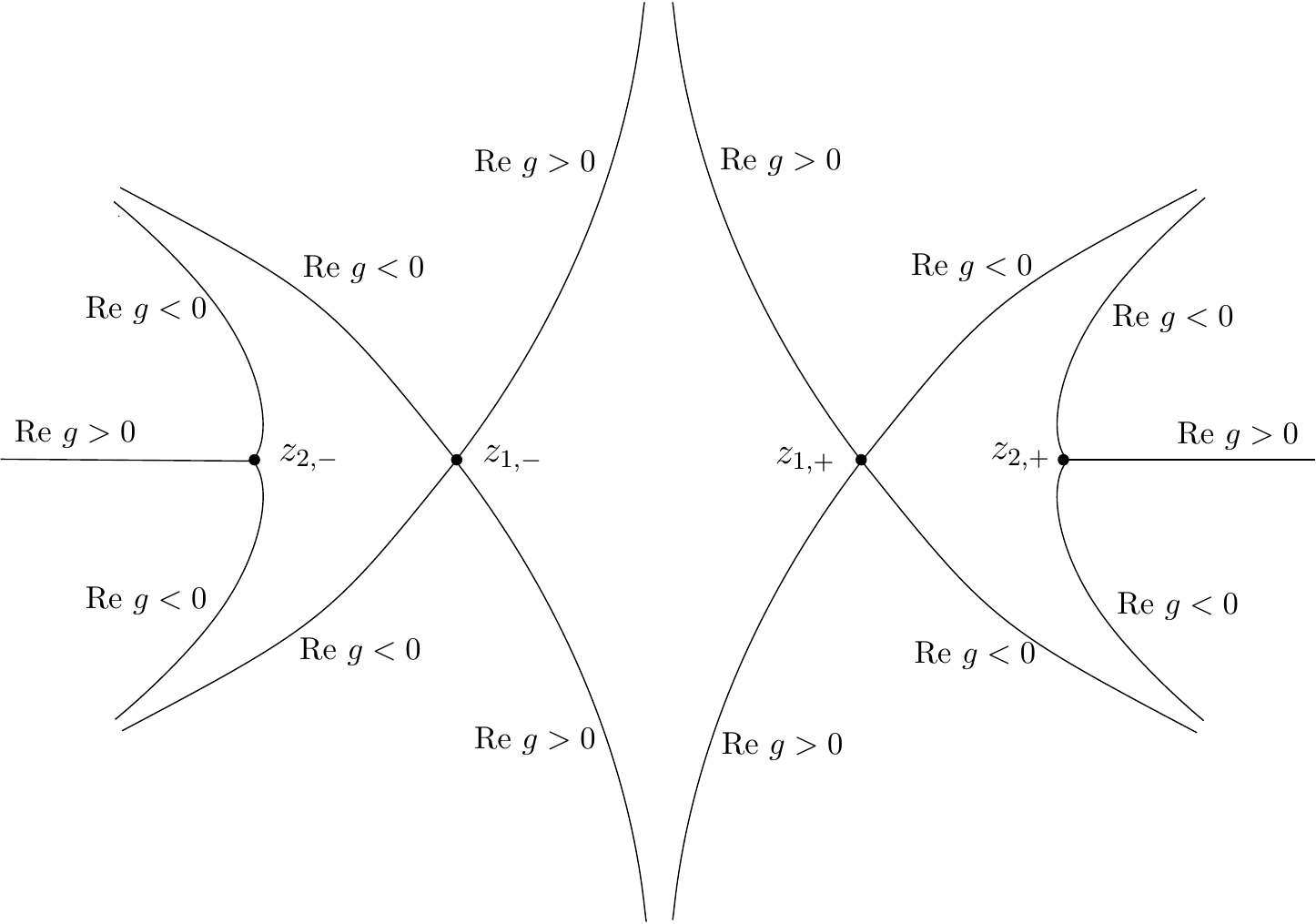}\\
  \caption{The anti-Stokes curves of the exponent $g(z)$}\label{ASC}
\end{figure}

\begin{figure}[H]
  \centering
  \includegraphics[width=11cm,height=4.95cm]{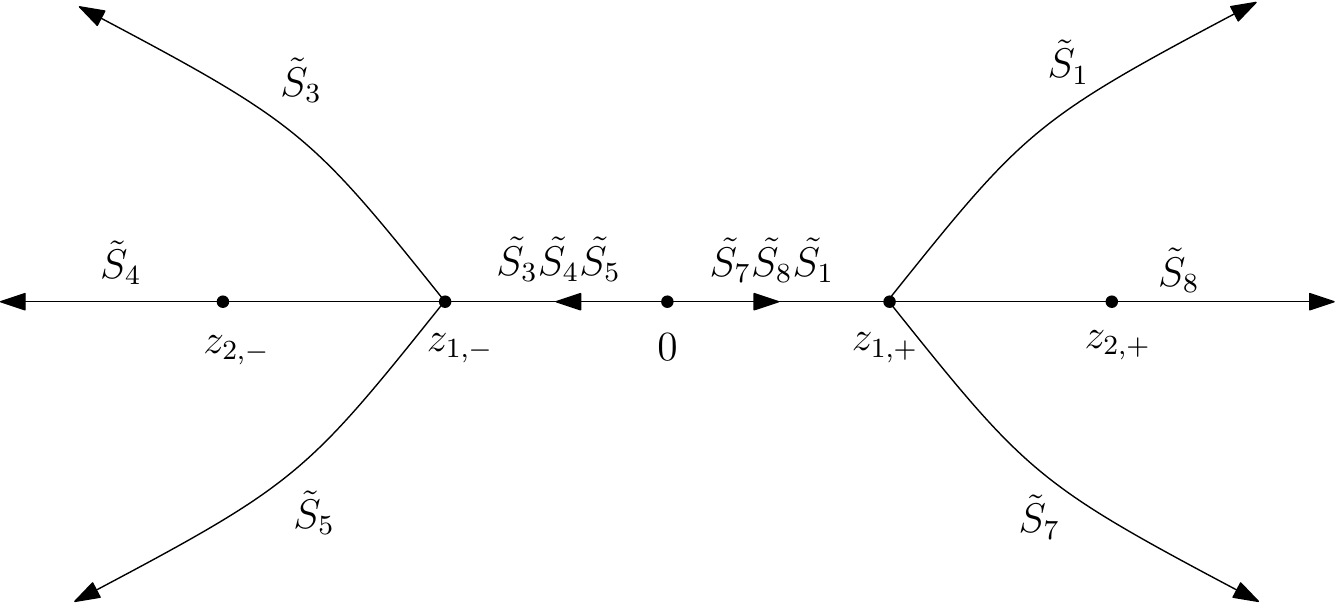}\\
  \caption{The first deformation of the jump curves of the RH problem} \label{Deforma1}
\end{figure}

Next, we transform the above RH problem into a  RH problem formulated on the anti-Stokes curves of $g(z)$, as depicted in Figure \ref{ASC}. To this end, first we note that the above RH problem for $U(z)$ can be rewritten as the RH problem posed on the curves shown in Figure \ref{Deforma1}, where we have used the notations $\widetilde{S}_k$, $k=1,3,4,5,7,8$ to denote the corresponding jump matrices $J_{U}(z)$.

It is now seen that the diagonal entries of jump matrices on $[z_{2,-},z_{2,+}]$ are highly oscillating for large $|x|$. To turn the oscillations into exponential decays on the anti-Stokes curves of $g(z)$, we  deform the segment $[z_{2,-},z_{2,+}]$ and therefore introduce the third transformation $U\rightarrow T$. This transformation is based on the following factorizations
\begin{align}\nonumber
\widetilde{S}_{4}&=\begin{pmatrix}e^{x^2(g_-(z)-g_+(z))} & 0\\ -s_0e^{2\pi i\alpha}z^{-2\alpha}e^{-x^2(g_+(z)+g_-(z))} & e^{x^2(g_+(z)-g_-(z))}\end{pmatrix}\\ \nonumber
&=\begin{pmatrix}1 & -s_0^{-1}e^{2\pi i\alpha}z^{2\alpha}e^{2x^2g_-(z)} \\ 0 & 1\end{pmatrix}
\begin{pmatrix}0 & s_0^{-1}|z|^{2\alpha} \\ -s_0|z|^{-2\alpha} & 0\end{pmatrix}
\begin{pmatrix}1 & -s^{-1}_0e^{-2\pi i\alpha}z^{2\alpha}e^{2x^2g_+(z)} \\ 0 & 1\end{pmatrix}\\
&=:\widetilde{S}_{U_1}\widetilde{S}_{P_-}\widetilde{S}_{U_2}, \label{decom1}
\end{align}
\begin{align} \nonumber
\widetilde{S}_{8}&=\begin{pmatrix}
e^{x^2(g_-(z)-g_+(z))} & 0\\ s_0|z|^{-2\alpha}e^{-x^2(g_+(z)+g_-(z))} & e^{x^2(g_+(z)-g_-(z))}
\end{pmatrix}\\ \nonumber
&=\begin{pmatrix}1 & s_0^{-1}e^{-4\pi i\alpha}(z^{2\alpha})_-e^{2x^2g_-(z)} \\ 0 & 1\end{pmatrix}
\begin{pmatrix}0 & -s_0^{-1}|z|^{2\alpha} \\ s_0|z|^{-2\alpha} & 0\end{pmatrix}
\begin{pmatrix}1 & s^{-1}_0(z^{2\alpha})_+e^{2x^2g_+(z)} \\ 0 & 1\end{pmatrix}\\
&=:\widetilde{S}_{U_3}\widetilde{S}_{P_+}\widetilde{S}_{U_4},\label{decom2}
\end{align}

\begin{align}\nonumber
(\widetilde{S}_3\widetilde{S}_4\widetilde{S}_5)^{-1}&=\begin{pmatrix}s_*
e^{x^2(g_-(z)-g_+(z))} & s_1(e^{-2\pi i\alpha}+s_*)z^{2\alpha}\\ s_0e^{2\pi i\alpha}z^{-2\alpha}& (1+s_0s_1e^{2\pi i\alpha})e^{x^2(g_+(z)-g_-(z))}\end{pmatrix}\\ \nonumber
&=\begin{pmatrix}1 & 0 \\ \frac{\overline{s}_*z^{-2\alpha}e^{-2x^2g_-(z)}}{s_1(e^{-2\pi i\alpha}+s_*)} & 1\end{pmatrix}
\begin{pmatrix}0 &  \frac{(|s_*|^2-1) |z|^{2\alpha} }{s_0}\\ \frac{s_0|z|^{-2\alpha}}{1-|s_*|^2 } & 0\end{pmatrix}
\begin{pmatrix}1 & 0 \\ \frac{s_*z^{-2\alpha}e^{-2x^2g_+(z)}}{s_1(e^{-2\pi i\alpha}+s_*)} & 1\end{pmatrix}\\ \label{decom3}
&=:\widetilde{S}_{L_1}\widetilde{S}_P\widetilde{S}_{L_2},
\end{align}
and
\begin{align} \nonumber
\widetilde{S}_7\widetilde{S}_8\widetilde{S}_1&=\begin{pmatrix}(1+s_0s_1e^{2\pi i\alpha})e^{x^2(g_-(z)-g_+(z))} & s_1(1+s_*e^{2\pi i\alpha})|z|^{2\alpha} \\ s_0|z|^{-2\alpha}& s_*e^{x^2(g_+(z)-g_-(z))}\end{pmatrix}\\ \nonumber
&=\begin{pmatrix}1 & 0 \\ \frac{s_*(z^{-2\alpha})_-e^{-2x^2g_-(z)}}{e^{-2\pi i\alpha}s_1(e^{-2\pi i\alpha}+s_*)} & 1\end{pmatrix}
\begin{pmatrix}0 &  \frac{(|s_*|^2-1) |z|^{2\alpha} }{s_0}\\ \frac{s_0|z|^{-2\alpha}}{1-|s_*|^2 } & 0\end{pmatrix}
\begin{pmatrix}1 & 0 \\ \frac{\overline{s}_*(z^{-2\alpha})_+e^{-2x^2g_+(z)}}{e^{2\pi i\alpha}s_1(e^{-2\pi i\alpha}+s_*)} & 1\end{pmatrix}\\ \label{decom4}
&=:\widetilde{S}_{L_3}\widetilde{S}_P\widetilde{S}_{L_4}.
\end{align}
It should be mentioned that in the above factorizations, we have used the property
$$g_+(z)+g_-(z)=0, \quad z\in [z_{2,-},z_{2,+}], $$
 and the complex conjugate relation
\begin{equation}\label{sstarbar}
\overline{s}_*=1+s_0s_1e^{2\pi i\alpha},
\end{equation}
which follows from \eqref{srela1}, \eqref{assum1} and \eqref{sstar}.

\begin{figure}[H]
  \centering
  \includegraphics[width=11cm,height=4.95cm]{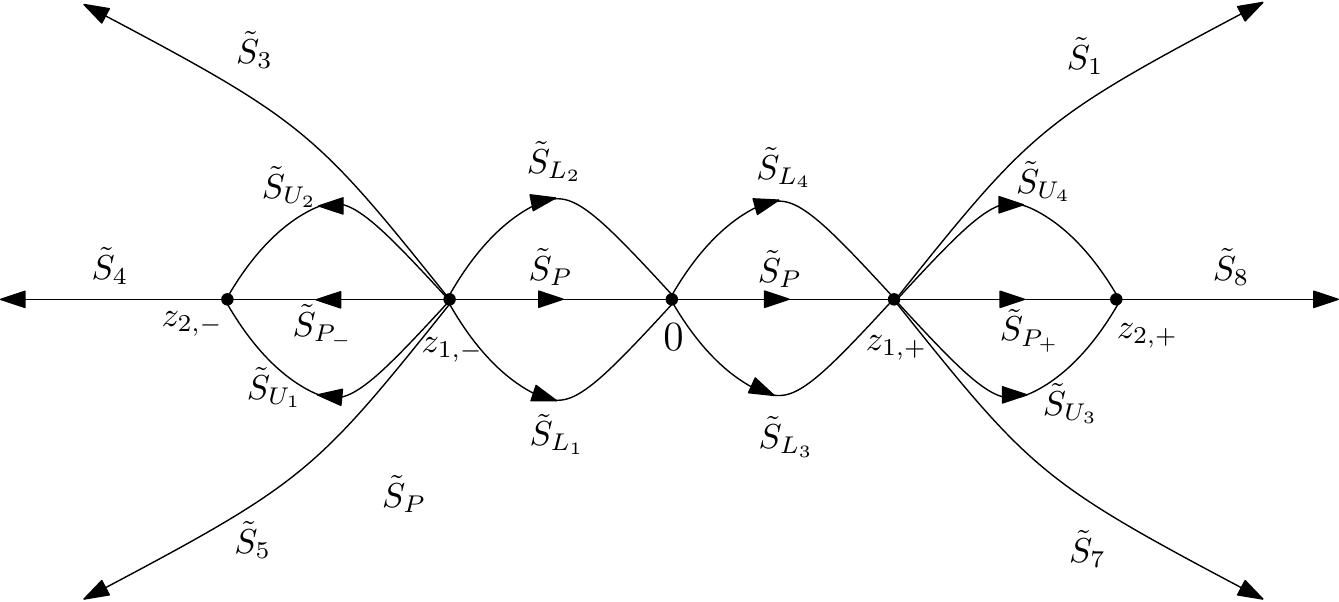}\\
  \caption{The second deformation of the jump curves of the RH problem} \label{Deforma2}
\end{figure}

Based on these matrix factorizations, we  obtain an equivalent RH problem formulated on the curves shown in Figure \ref{Deforma2}, where we have used the same notations to stand for the analytic extensions of jump matrices $\widetilde{S}_{U_k}(z)$ and $\widetilde{S}_{L_k}(z)$, $k=1,2,3,4$. In the next step,  to deform the jump curves into the anti-Stokes lines of $g(z)$ as shown in Figure \ref{ASC}, we  blow up the four lens.  As a consequence, we arrive at the following RH problem for $T(z)$.

\begin{figure}[H]
  \centering
  \includegraphics[width=11cm,height=7.7cm]{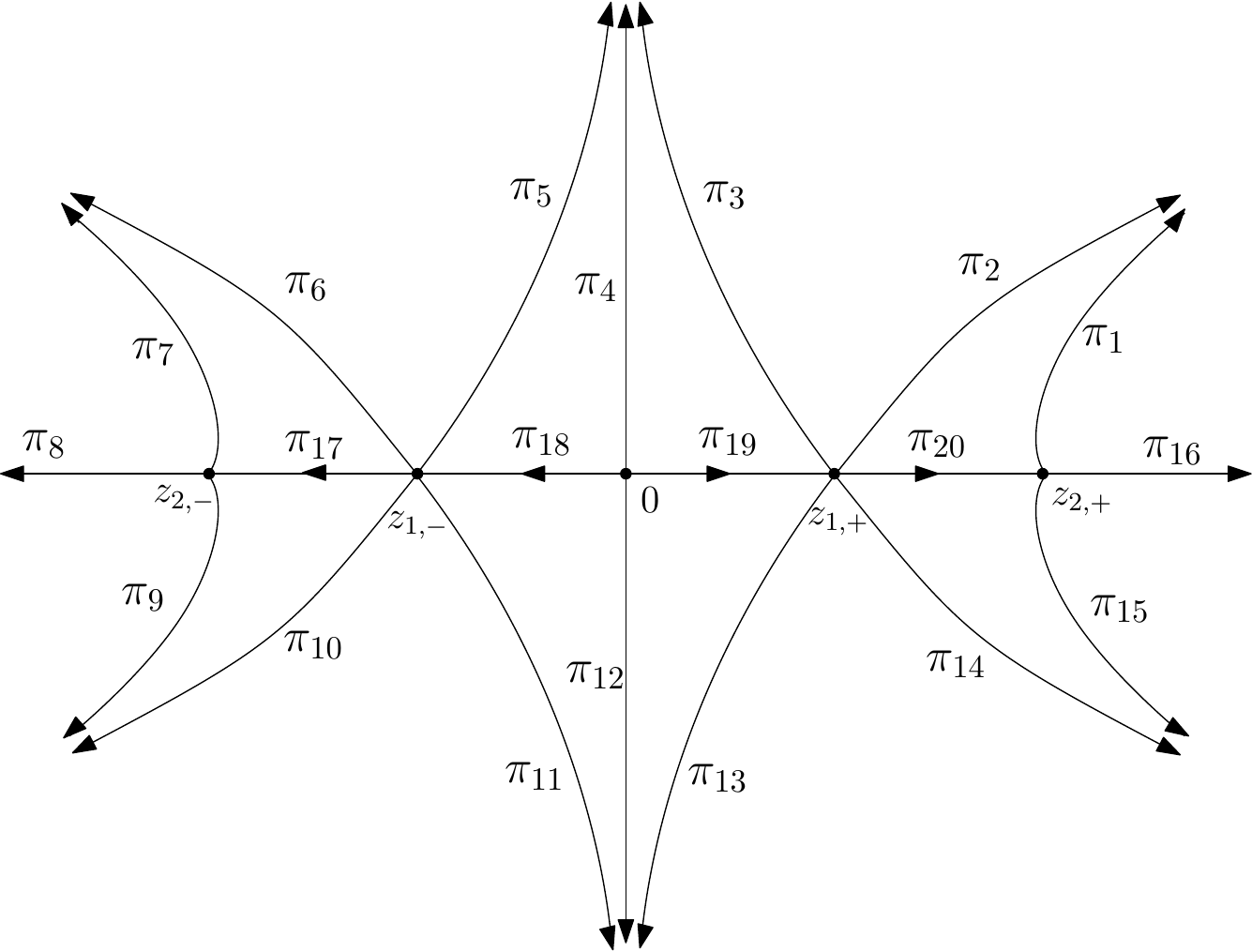}\\
  \caption{The final jump curves $\Sigma_T$ of the RH problem for $T(z)$} \label{Deforma3}
\end{figure}

\subsection*{RH problem for $T(z)$}
\begin{description}
  \item{(1)} $T(z)$ is analytic for $z\in \mathbb{C}\setminus\Sigma_T$, where $\Sigma_T=\bigcup^{20}_{k=1}\pi_k$ is depicted in Figure \ref{Deforma3}.
  \item{(2)} $T(z)$ satisfies the jump relations $T_+(z)=T_-(z)J_k(z)$ for $z\in\pi_k$, where
    \begin{align*}
  &J_1(z)=\begin{pmatrix}1 & -s_0^{-1}z^{2\alpha}e^{2x^2g(z)}\\ 0 & 1 \end{pmatrix},
   &&
  J_2(z)=\begin{pmatrix}1 & s_*s_0^{-1}z^{2\alpha}e^{2x^2g(z)}\\ 0 & 1 \end{pmatrix},\\
  &J_3(z)=\begin{pmatrix}1 & 0\\ -\frac{\bar s_*z^{-2\alpha}e^{-2x^2g(z)}}
  {e^{2\pi i\alpha}s_1(e^{-2\pi i\alpha}+s_*)} & 1 \end{pmatrix}, &&
  J_4(z)=\begin{pmatrix}1 & 0\\ \frac{(e^{-2\pi i\alpha}-1)z^{-2\alpha}e^{-2x^2g(z)}}
  {s_1(e^{-2\pi i\alpha}+s_*)} & 1 \end{pmatrix},\\
  &J_5(z)=\begin{pmatrix}1 & 0\\ \frac{s_*z^{-2\alpha}e^{-2x^2g(z)}}
  {s_1(e^{-2\pi i\alpha}+s_*)} & 1 \end{pmatrix}
  , &&
  J_6(z)=\begin{pmatrix}1 & -\frac{\bar s_*z^{2\alpha}
  e^{2x^2g(z)}}{s_0e^{2\pi i\alpha}} \\ 0 & 1 \end{pmatrix},\\
  &J_7(z)=J_9(z)=\begin{pmatrix}1 & \frac{s_0^{-1}
  z^{2\alpha}e^{2x^2g(z)}}{e^{2\pi i\alpha}} \\ 0 & 1 \end{pmatrix},&&
  J_8(z)=\begin{pmatrix}1 & 0\\ -s_0e^{2\pi i\alpha}z^{-2\alpha}e^{-2x^2g(z)} & 1 \end{pmatrix},\\
  &J_{10}(z)=\begin{pmatrix}1 & -\frac{s_*z^{2\alpha}
  e^{2x^2g(z)}}{s_0e^{2\pi i\alpha}}\\ 0 & 1 \end{pmatrix}
  ,&&
  J_{11}(z)=\begin{pmatrix}1 & 0\\ \frac{\bar s_*z^{-2\alpha}e^{-2x^2g(z)}}
  {s_1(e^{-2\pi i\alpha}+s_*)} & 1 \end{pmatrix},\\
  &J_{12}(z)=\begin{pmatrix}1 & 0\\ \frac{(e^{2\pi i\alpha}-1)z^{-2\alpha}e^{-2x^2g(z)}}
  {s_1(e^{-2\pi i\alpha}+s_*)} & 1 \end{pmatrix},&&
   J_{13}(z)=\begin{pmatrix}1 & 0\\ -\frac{s_*e^{2\pi i\alpha}z^{-2\alpha}e^{-2x^2g(z)}}
  {s_1(e^{-2\pi i\alpha}+s_*)} & 1 \end{pmatrix},\\
  &J_{14}(z)=\begin{pmatrix}1 & \frac{\bar{s}_*z^{2\alpha}e^{2x^2g(z)}}{s_0e^{4\pi i\alpha}}\\ 0 & 1 \end{pmatrix},
  &&
  J_{15}(z)=\begin{pmatrix}1 & -\frac{z^{2\alpha}e^{2x^2g(z)}}{s_0e^{4\pi i\alpha}}\\ 0 & 1 \end{pmatrix},\\
  &J_{16}(z)=\begin{pmatrix}1 & 0\\ \frac{s_0e^{-2x^2g(z)}}{|z|^{2\alpha}} & 1 \end{pmatrix},&&
  J_{17}(z)=J_{20}(z)^{-1}=\begin{pmatrix}0 & s_0^{-1}|z|^{2\alpha}\\ -\frac{s_0}{|z|^{2\alpha}} & 0 \end{pmatrix},
  \end{align*}
  and
  \begin{equation*}
  J_{19}(z)=J_{18}(z)^{-1}=\begin{pmatrix}0 & \begin{smallmatrix}s_1(1+s_*e^{2\pi i\alpha})|z|^{2\alpha}\end{smallmatrix}\\ \frac{-|z|^{-2\alpha}}{s_1(1+s_*e^{2\pi i\alpha})} & 0\end{pmatrix}.
  \end{equation*}
  \item{(3)} $T(z)=I+O(z^{-1})$ as $z\rightarrow \infty$.
  \item{(4)} $T(z)$ has the following asymptotic behavior near the origin
  \begin{equation}\label{T0}
T(z)=T_0(z)z^{\alpha\sigma_{3}}
E_0S_1 z^{-\alpha\sigma_3}e^{-x^2g(z)\sigma_{3}}J_3(z),\end{equation}
where $\arg z\in (0, {\pi}/{2})$ and $T_0(z)$ is analytic in a neighborhood of $z=0$. The behavior of $T(z)$ in other regions  is determined by \eqref{T0} and  the jump relations satisfied by $T(z)$.
\end{description}

Using the lower and upper triangular structure of the jump matrices, the sign of $\Re g(z)$ on the anti-Stokes curves (cf. Figure \ref{ASC}) and the property that $\Re g(z)>0$ on the imaginary axis, it follows that the jump matrices for $T$ tend to the identity matrix exponentially fast as $x\rightarrow -\infty$, except the ones on the segment $[z_{2,-},z_{2,+}]$. In the next subsections, we shall construct the global parametrix with jumps on the segment $[z_{2,-},z_{2,+}]$ and the local parametrices near the saddle points  $z_{1,\pm}=\pm\sqrt{2/3}$, $z_{2,\pm}=\pm\sqrt{8/3}$ and the origin.

\subsection{Global parametrix on $[z_{2,-},z_{2,+}]$}
Orienting the line segment $[z_{2,-},z_{2,+}]$ rightward, we are now in a position  to solve the following RH problem for a $2\times 2$ matrix-valued function $P^{(\infty)}(z)$.
\subsection*{RH problem for $P^{(\infty)}(z)$}
\begin{description}
  \item{(1)} $P^{(\infty)}(z)$ is analytic for $z\in \mathbb{C}\setminus [z_{2,-},z_{2,+}]$.
  \item{(2)} $P^{(\infty)}(z)$ satisfies the jump relations
  \begin{equation}\label{Pinftyjump}
  P^{(\infty)}_{+}(z)=P^{(\infty)}_{-}(z)J_{\infty}(z),
  \end{equation}
  where
  $$
  J_{\infty}(z)=\left\{\begin{aligned}
  &\begin{pmatrix}0 & -s^{-1}_0|z|^{2\alpha} \\ s_0|z| ^{-2\alpha} & 0\end{pmatrix},& z&\in[z_{2,-},z_{1,-}], \\
  &\begin{pmatrix}0 & s^{-1}_0\left(|s_*|^2 -1\right )|z|^{2\alpha}\\  s_0\left(1-|s_*|^2\right )^{-1}|z|^{-2\alpha} & 0\end{pmatrix}, & z&\in[z_{1,-},z_{1,+}],\\
  &\begin{pmatrix}0 & -s^{-1}_0|z|^{2\alpha} \\ s_0|z|^{-2\alpha} & 0 \end{pmatrix}, & z&\in[z_{1,+},z_{2,+}].
  \end{aligned}\right.
  $$
  \item{(3)} $P^{(\infty)}(z)$ have at most singularities of order $\frac{3}{2}$ at $z=z_{1,\pm}$, respectively.
  \item{(4)}  As $z\rightarrow \infty$, we have $P^{(\infty)}(z)=I+O(z^{-1})$.
\end{description}

A solution of the above RH problem is given by
\begin{equation}\label{Pinfty}
P^{(\infty)}(z)=H(z)s_0^{-\frac{\sigma_3}{2}}D^{-\sigma_3}_{\infty}X(z)
D(z)^{\sigma_3}s_0^{\frac{\sigma_3}{2}},
\end{equation}
where $X(z)$ is given by
\begin{equation}\label{X(z)}
X(z)=\frac{1}{2}
\begin{pmatrix}\omega+\omega^{-1} & i(\omega-\omega^{-1}) \\ -i(\omega-\omega^{-1}) & \omega+\omega^{-1}\end{pmatrix},\quad \omega=\omega(z)=\left(\frac{z-\sqrt{\frac{8}{3}}}
{z+\sqrt{\frac{8}{3}}}\right)^{ {1}/{4}},
\end{equation}
the Szeg\"{o} function
\begin{align}\label{D}
D(z)=\left(\frac{3}{8}\right)^{\frac{\alpha}{2}}
\left(z+\left(z^2-\frac{8}{3}\right)
^{\frac{1}{2}}\right)^{\alpha}z^{-\alpha}\left(\frac{\left((2-\sqrt{3})z
-i\left(z^2-\frac{8}{3}\right)^{\frac{1}{2}}
\right)^2-\frac{8}{3}}
{\left((2-\sqrt{3})z+i\left(z^2-\frac{8}{3}\right)^{\frac{1}{2}}\right)^2
-\frac{8}{3}}\right)^{\nu},
\end{align}
and
\begin{equation}\label{Dinfty}
D_{\infty}=\lim\limits_{z\rightarrow\infty}D(z)=2^{-\frac{\alpha}{2}}
3^{\frac{\alpha}{2}}e^{\frac{\pi i\nu}{3}}.
\end{equation}
The branches of the functions in \eqref{X(z)} and \eqref{D} are chosen such that
$$\arg z\in(-\pi,\pi),\quad \arg\left(z\pm\textstyle\sqrt{\frac{8}{3}}\right)\in(-\pi,\pi),\quad \arg \left(\textstyle z+\left(z^2-\frac{8}{3}\right)^{\frac{1}{2}}\right)\in(-\pi,\pi)
$$
and
$$
\arg\left(\frac{\left((2-\sqrt{3})z-i\left(z^2-\frac{8}{3}\right)^{\frac{1}{2}}
\right)^2-\frac{8}{3}}
{\left((2-\sqrt{3})z+i\left(z^2-\frac{8}{3}\right)^{\frac{1}{2}}\right)^2
-\frac{8}{3}}\right)\in(-\pi,\pi).
$$
The  exponent $\nu$ in \eqref{D} is defined by
\begin{equation}\label{nu}
\nu=-\frac{1}{2\pi i}\ln(|s_*|^2-1)-\frac{1}{2}=:\nu_0-\frac{1}{2}\quad \mathrm{with}\quad\nu_0\in i\mathbb{R},
\end{equation}
noting that   $|s_*|>1$ in the case we are considering.

The factor
$H(z)$ in \eqref{Pinfty} is brought in
  to meet  the matching conditions \eqref{P1+matching} and \eqref{P1-matching} below. We seek for a  meromorphic function of the    form
\begin{equation}\label{H}
H(z)=I+\frac{A}{z-\sqrt{\frac{2}{3}}}+\frac{B}{z+\sqrt{\frac{2}{3}}}
\end{equation}
with
\begin{equation}\label{Hdet}
\det H(z)=1,
\end{equation}
where the constant matrices $A$ and $B$   are to be determined. Moreover, since $P^{(\infty)}(z)$ satisfies the  symmetric relation $P^{(\infty)}(z)=\sigma_3P^{(\infty)}(-z)\sigma_3$, we also require that $H(z)=\sigma_3H(-z)\sigma_3$. Therefore,  $A$ and $B$ are subject to the constraint \begin{equation}\label{symrela}
A=-\sigma_3 B\sigma_3.
\end{equation}

\begin{rem}\label{rem:singular-parametrix}
$H(z)$ in \eqref{H} brings extra poles $z=\pm \sqrt{\frac 2 3}$ to the global parametrix \eqref{Pinfty} for $P^{(\infty)}(z)$.
Such obstacles also arose  in   deriving singular asymptotics for the PII
transcendents \cite{BI,Hu}.
  In \cite{BI}, Bothner and Its developed a certain {\it dressing}  technique to transform the RH problem to another one without poles.  Similar matching technique was used earlier in  \cite{ZZ}  to derive a uniform asymptotic approximation of the Pollaczek polynomials, and then in \cite{ZXZ} for an asymptotic study  of a system of Szeg\H{o} class polynomials.
\end{rem}

\subsection{Local parametrices near $z_{1,\pm}$ } %and determinations of $A$ and $B$}
In this subsection, we construct two  parametrices $P^{(1,\pm)}(z)$ satisfying the same jump conditions as $T(z)$ on the contours $\Sigma_T$ (see Figure \ref{Deforma3}) respectively  in the neighborhoods $U(z_{1,\pm},\delta)$ of the saddle points $z_{1,\pm}=\pm \sqrt{2/3}$ and matching with $P^{(\infty)}(z)$ on the boundaries $\partial U(z_{1,\pm},\delta)$.

\subsection*{RH problem for $P^{(1,+)}(z)$}
\begin{description}
\item{(1)} $P^{(1,+)}(z)$ is analytic for  $z\in U(z_{1,+},\delta)\setminus\Sigma_T$, where $U(z_{1,+},\delta)=\{z\in \mathbb{C}:|z-z_{1,+}| <\delta\}$.
\item{(2)} $P^{(1,+)}(z)$ shares the same jump conditions  as $T(z)$ on $U(z_{1,+},\delta)\cap\Sigma_T$.
\item{(3)} On the boundary of the disc $\partial U(z_{1,+},\delta)=\{z\in \mathbb{C}:|z-z_{1,+}|=\delta\}$,
\begin{equation}\label{P1+matching}
P^{(1,+)}(z)=\left(I+O(|x|^{-2})\right)P^{(\infty)}(z),\quad \mathrm{as}\quad x\rightarrow -\infty.
\end{equation}
\end{description}
In order to construct a solution to the above RH problem, first we define the conformal mapping
\begin{equation}\label{varphi}
\varphi(z)=\left\{\begin{aligned}
&2\sqrt{-\textstyle\frac{\sqrt{3}i}{6}-g(z)},\quad& \Im z>0,\\
&2\sqrt{-\textstyle\frac{\sqrt{3}i}{6}+g(z)},\quad& \Im z<0,
\end{aligned}\right.
\end{equation}
where the branches of the square roots are specified choosing
\begin{equation}\label{varphibehavior}
\varphi(z)=e^{-\frac{\pi i}{4}}2\cdot3^{-\frac{1}{4}}\textstyle \left(z-\sqrt{\frac{2}{3}}\right)
\left(1+o(1)\right),\quad \mathrm{as}\quad z\rightarrow\sqrt{\frac{2}{3}}.
\end{equation}
Let $\Phi^{\mathrm{(PC)}}$ be the  parabolic cylinder parametrix  given in Appendix \ref{PCP} with the parameter
$\nu$ defined by  \eqref{nu}.
Then, the solution to above RH problem can be  constructed as follows:
\begin{align}\label{P1+}
P^{(1,+)}(z)=E^{(1,+)}(z)\Phi^{\mathrm{(PC)}}\left(|x|\varphi(z)\right)
\left(\frac{s_*}{h_0}\right)^{\frac{\sigma_3}{2}}
\left\{\begin{aligned}
&s_0^{-\frac{\sigma_3}{2}}(-\sigma_{1})z^{-\alpha\sigma_3}
e^{-x^2g(z)\sigma_3},& \Im z>0,\\
&s_0^{\frac{\sigma_3}{2}}\sigma_3e^{2\pi i(\alpha+\nu)\sigma_3}
z^{-\alpha\sigma_3}e^{-x^2g(z)\sigma_3},& \Im z<0,
\end{aligned}\right.
\end{align}
where $h_{0}$ is defined in  \eqref{h0} and $E^{(1,+)}(z)$ is given by
\begin{equation}\label{E1+}
E^{(1,+)}(z)=W^{(+)}(z)\left(\frac{s_*}{h_0}\right)^{-\frac{\sigma_3}{2}}
|x|^{\nu\sigma_3}e^{i\frac{\sqrt{3}}{6}x^2
\sigma_{3}}\begin{pmatrix}1 & 0 \\ -\frac{1}{|x|\varphi(z)} & 1\end{pmatrix}2^{-\frac{\sigma_3}{2}}\begin{pmatrix}|x|\varphi(z) & 1 \\ 1 & 0\end{pmatrix},
\end{equation}
with
\begin{equation}\label{W+}
W^{(+)}(z)=\left\{\begin{aligned}
&P^{(\infty)}(z)z^{\alpha\sigma_3}
(-\sigma_1)s_0^{\frac{1}{2}\sigma_3}\varphi(z)^{\nu\sigma_3},
&\Im z&>0,\\
&P^{(\infty)}(z)z^{\alpha\sigma_3}e^{-2\pi i(\alpha+\nu)\sigma_3}\sigma_3s_0^{-\frac{1}{2}\sigma_3}
\varphi(z)^{\nu\sigma_3}, & \Im z&<0.
\end{aligned}\right.
\end{equation}
Here, the branch of the function $\varphi(z)^{\nu}$ is chosen  by requiring  $\arg\varphi(z)\in(0,2\pi)$.  This leads to the jump relations
\begin{equation}\label{tildezetajump}
\left\{\begin{aligned}
&\left(\varphi(z)^{\nu}\right)_+=
\left(\varphi(z)^{\nu}\right)_-e^{2\pi i\nu}, \quad &z&\in[z_{1,+},z_{2,+}],\\
&\left(\varphi(z)^{\nu}\right)_+
=\left(\varphi(z)^{\nu}\right)_-, \quad &z&\in[z_{2,-},z_{1,+}].
\end{aligned}\right.
\end{equation}
Using \eqref{Pinftyjump} and \eqref{tildezetajump}, it is readily verified  that $E^{(1,+)}(z)$ is holomorphic in the deleted neighborhood $U(z_{1,+},\delta)\setminus\{z_{1,+}\}$. The analyticity of  $E^{(1,+)}(z)$ at the isolated point $z_{1,+}=\sqrt{\frac{2}{3}}$ will be guaranteed by a proper choice of the constant matrices $A$ and $B$  in \eqref{H}.

Indeed, by computing the Laurent expansion of $E^{(1,+)}(z)$ at $z_{1,+}=\sqrt{\frac{2}{3}}$ using \eqref{Pinfty}, \eqref{X(z)}, \eqref{D} and \eqref{varphibehavior}, we have
\begin{equation}\label{mE1+}
E^{(1,+)}(z)=\left(\frac{A}{\tau}+I+\sqrt{\frac{3}{8}}B+O(\tau)\right)
\left(I-\frac{c}{\sqrt 3 \tau}\begin{pmatrix}1 & -e^{-\frac{\pi i} 6}s_0^{-1}D_{\infty}^{-2} \\ e^{ \frac{\pi i} 6}s_0D_{\infty}^2 & -1\end{pmatrix}\right)M_1(z),
\end{equation}
where $\tau=z-z_{1,+}$, $M_1(z)$ is analytic near $z=z_{1,+}$ and $c=c(x)$ is given by
\begin{equation}\label{c}
c=-\frac{i\sqrt{6}e^{i\phi}}{2+e^{i\phi}},\quad \phi=-\frac{\sqrt{3}}{3}x^{2}+i\nu_0\ln \left(2 \sqrt{3} x^{2}\right)
+\frac{2\pi\alpha}{3}+\arg\Gamma\left(\nu_0+ \frac{1}{2}\right)+\arg s_{*},
\end{equation}
with $\nu_0$  given in \eqref{nu}.

To ensure that $ E^{(1,+)}(z)$ is holomorphic at $z_{1,+}=\sqrt{\frac{2}{3}}$, it is seen from \eqref{mE1+} that $A$ and $B$ must fulfill the following algebraic equations \begin{align}\label{equat1}
A&=
\frac{c}{\sqrt 3 }\left(I+\sqrt{\frac{3}{8}}B\right)\begin{pmatrix}1 & -e^{-\frac{\pi i} 6}s_0^{-1}D_{\infty}^{-2} \\ e^{ \frac{\pi i} 6}s_0D_{\infty}^2 & -1\end{pmatrix}
,\\
0&=A\begin{pmatrix}1 & -e^{-\frac{\pi i} 6}s_0^{-1}D_{\infty}^{-2} \\ e^{ \frac{\pi i} 6}s_0D_{\infty}^2 & -1\end{pmatrix}.\label{equat2}
\end{align}
Equation \eqref{equat2} follows directly from \eqref{equat1}
since the second matrix on the right-hand side of \eqref{equat1} is nilpotent. A combination of  the  equation  \eqref{equat1} with the symmetric  condition  \eqref{symrela} gives us the explicit expressions of $A$ and $B$, namely
\begin{equation}\label{repAB}
A=\sqrt{\frac 2 3} \begin{pmatrix} \frac{  c}{ c+\sqrt 2}   &
-e^{-\frac{\pi i} 6} \frac{ c\, s_0^{-1}D^{-2}_{\infty}}
{ \sqrt 2+c }   \\[.3cm]
 e^{ \frac{\pi i} 6} \frac{ c\, s_0 D^{ 2}_{\infty}}
{ \sqrt 2-c }   &  \frac{c}{  c-\sqrt{2} }
\end{pmatrix},\quad
B=\sqrt{\frac 2 3} \begin{pmatrix}    -\frac{  c}{ c+\sqrt 2}  &
-e^{-\frac{\pi i} 6} \frac{ c\, s_0^{-1}D^{-2}_{\infty}}
{ \sqrt 2+c }   \\[.3cm]
e^{ \frac{\pi i} 6} \frac{ c\, s_0 D^{ 2}_{\infty}}
{ \sqrt 2-c }  &  -\frac{c}{  c-\sqrt{2} }
\end{pmatrix}.
\end{equation}
Having determined $A$ and $B$, straightforward verification shows that   the determinant condition \eqref{Hdet} holds.

It should be mentioned that we assume in \eqref{repAB} that $x$ lies outside of the zero sets of the functions $\sqrt{2}\pm   c(x)$, which consist of two sequences of points $\{x_n\}$ and  $\{y_n\}$ for  $n\in \mathbb{N}$, defined respectively by the equations
\begin{equation}\label{pole1}
-\frac{\sqrt{3}}{3}x_n^{2}+i\nu_0\ln \left(2 \sqrt{3} x_n^{2}\right)
+\frac{2\pi\alpha}{3} +\arg \Gamma\left(\nu_0+ \frac{1}{2}\right)+\arg s_{*}+\frac{2\pi}{3}+2n\pi=0,
\end{equation}
and
\begin{equation}\label{pole2}
-\frac{\sqrt{3}}{3}y_n^{2}+i\nu_0\ln \left(2 \sqrt{3} y_n^{2}\right)
+\frac{2\pi\alpha}{3} +\arg \Gamma\left(\nu_0+ \frac{1}{2}\right)+\arg s_{*}-\frac{2\pi}{3}+2n\pi=0.
\end{equation}
As we will see later, these points are the singularities appeared in the leading term of the asymptotic formula \eqref{qasymp-infty}. More precisely,  $\{x_n\}$ and  $\{y_n\}$ are approximate to the simple poles of $q(x; \kappa )$  on the negative real axis, such that
$x_n\sim a_n^+$ and $y_n\sim a_n^-$ as $n\to\infty$; see equation \eqref{poles} in Corollary \ref{cor}.

Finally, a combination of \eqref{Pinfty}, \eqref{H}, \eqref{P1+} and \eqref{PCAsyatinfty} gives us the matching condition \eqref{P1+matching}.

\subsection*{RH problem for $P^{(1,-)}(z)$}
\begin{description}
\item{(1)} $P^{(1,-)}(z)$ is analytic for   $z\in U(z_{1,-},\delta)\setminus\Sigma_T$, where $U(z_{1,-},\delta)=\{z\in \mathbb{C}:|z-z_{1,-}|<\delta\}$.
\item{(2)} $P^{(1,-)}(z)$ satisfies the same jump conditions as $T(z)$ on $U(z_{1,-},\delta)\cap\Sigma_T$.
\item{(3)} On the boundary $\partial U(z_{1,-},\delta)=\{z\in \mathbb{C}:|z-z_{1,-}|=\delta\}$,
\begin{equation}\label{P1-matching}
P^{(1,-)}(z)=\left(I+O\left(|x|^{-2}\right)\right)P^{(\infty)}(z),\quad \mathrm{as}\quad x\rightarrow -\infty.
\end{equation}
\end{description}
Similar to the construction of $P^{(1,+)}(z)$, we introduce a conformal mapping
\begin{equation}\label{zeta}
\zeta(z)=\left\{\begin{aligned}
&2\sqrt{-\textstyle\frac{\sqrt{3}i}{6}+g(z)},\quad& \Im z>0,\\
&2\sqrt{-\textstyle\frac{\sqrt{3}i}{6}-g(z)},\quad& \Im z<0,
\end{aligned}\right.
\end{equation}
where the branches of the square roots are chosen such that
\begin{equation}\label{zetabeha}
\zeta(z)=e^{\frac{3\pi i}{4}}2\cdot3^{-\frac{1}{4}}\textstyle \left(z+\sqrt{\frac{2}{3}}\;\right)
\left(1+o(1)\right),\quad \mathrm{as}\quad z\rightarrow-\sqrt{\frac{2}{3}}.
\end{equation}
The solution to the above RH problem can be constructed in terms of the parabolic cylinder function as follows:
\begin{align}\label{P1-}
&P^{(1,-)}(z)=E^{(1,-)}(z)\Phi^{\mathrm{(PC)}}\left(|x|\zeta(z)\right)
\left(\frac{s_*}{h_0}\right)^{\frac{\sigma_3}{2}}\left\{\begin{aligned}
&e^{2\pi i\nu\sigma_3}\left(s_0e^{2\pi i\alpha}\right)^{\frac{\sigma_3}{2}}z^{-\alpha\sigma_3}e^{-x^2g(z)\sigma_3},& \Im z>0,\\
&\left(s_0e^{2\pi i\alpha}\right)^{-\frac{\sigma_3}{2}}
\sigma_{3}\sigma_{1}z^{-\alpha\sigma_3}e^{-x^2g(z)\sigma_3},
& \Im z<0,\end{aligned}\right.
\end{align}
where $\Phi^{\mathrm{(PC)}}$ is the  parabolic cylinder parametrix given in  Appendix \ref{PCP}, $h_{0}$ and $\nu$  are  defined in \eqref{h0} and \eqref{nu}, respectively.  Here,  $E^{(1,-)}(z)$ is given by
\begin{equation}\label{E1-}
E^{(1,-)}(z)=W^{(-)}(z)\left(\frac{s_*}{h_0}\right)^{-\frac{\sigma_3}{2}}
|x|^{\nu\sigma_3}e^{\frac{i\sqrt{3}x^2}{6}
\sigma_{3}}\begin{pmatrix}1 & 0 \\ -\frac{1}{|x|\zeta(z)} & 1\end{pmatrix}2^{-\frac{\sigma_3}{2}}\begin{pmatrix}|x|\zeta(z) & 1 \\ 1 & 0\end{pmatrix}
\end{equation}
with
\begin{equation}\label{W-}
W^{(-)}(z)=\left\{\begin{aligned}
&P^{(\infty)}(z)z^{\alpha\sigma_3}\left(s_0e^{2\pi i\alpha}\right)^{-\frac{\sigma_3}{2}}
e^{-2\pi i\nu\sigma_3}\zeta(z)^{\nu\sigma_3},\ &\Im z&>0,\\
&P^{(\infty)}(z)z^{\alpha\sigma_3}\sigma_1\sigma_3\left(s_0e^{2\pi i\alpha}\right)^{\frac{\sigma_3}{2}}\zeta(z)^{\nu\sigma_3}, \ &\Im z&<0.
\end{aligned}\right.
\end{equation}
The branch of the function $\zeta(z)^{\nu}$ is chosen such that $\arg\zeta(z)\in(-\pi,\pi)$. This implies that
\begin{equation}\label{zetajump}
\left\{\begin{aligned}
&\left(\zeta(z)^{\nu}\right)_+=\left(\zeta(z)^{\nu}\right)_-e^{2\pi i\nu}, \quad &z&\in[z_{2,-},z_{1,-}],\\
&\left(\zeta(z)^{\nu}\right)_+=\left(\zeta(z)^{\nu}\right)_-, \quad &z&\in[z_{1,-},z_{2,+}].
\end{aligned}\right.
\end{equation}
Using the jump relations \eqref{Pinftyjump} and \eqref{zetajump}, it is straightforward to check that $W^{(-)}(z)$ is holomorphic in $U(z_{1,-},\delta)$.
Furthermore, combining \eqref{Pinfty}, \eqref{H}, \eqref{P1-} with the asymptotic behavior \eqref{PCAsyatinfty}, we obtain the matching condition \eqref{P1-matching}.

\subsection{Local parametrices near saddle points $z_{2,\pm}$}
In this subsection, we seek two parametrices $P^{(2,\pm)}(z)$ satisfying the same jump conditions as $T(z)$ on  the  curves  $\Sigma_T$ (see Figure \ref{Deforma3}) in the neighbourhoods $U(z_{2,\pm},\delta)$ of the saddle points $z_{2,\pm}=\pm\sqrt{8/3}$, matching with $P^{(\infty)}(z)$ on the boundaries $\partial U(z_{2,\pm},\delta)$.
\subsection*{RH problem for $P^{(2,+)}(z)$}
\begin{description}
\item{(1)} $P^{(2,+)}(z)$ is analytic for $z\in U(z_{2,+},\delta)\setminus\Sigma_T$, where $U(z_{2,+},\delta)=\{z\in \mathbb{C}:|z-z_{2,+}|<\delta\}$.
\item{(2)} $P^{(2,+)}(z)$ satisfies the same jump conditions as $T(z)$ on $U(z_{2,+},\delta)\cap\Sigma_T$.
\item{(3)} On the boundary $\partial U(z_{2,+},\delta)=\{z\in \mathbb{C}:|z-z_{2,+}|=\delta\}$, we have
\begin{equation}\label{P2+matching}
P^{(2,+)}(z)=\left(I+O\left(|x|^{-2}\right)\right)P^{(\infty)}(z),\quad \mathrm{as}\quad x\rightarrow -\infty.
\end{equation}
\end{description}
To find a solution to the above RH problem, we  define the conformal mapping
\begin{equation}\label{eta}
\eta(z)=\left(\frac{3}{2}g(z)\right)^{\frac{2}{3}},
\end{equation}
where the branch is chosen such that
\begin{equation}\label{etabeha}
\eta(z)=2^{\frac{5}{6}}3^{-\frac{1}{6}}
\textstyle\left(z-\sqrt{\frac{8}{3}}\right)
\left(1+o(1)\right),\quad \mathrm{as} \quad z\rightarrow \sqrt{\frac{8}{3}}.
\end{equation}
Then, the solution to the above RH problem can be explicitly constructed in terms of the Airy function
\begin{equation}\label{P2+}
P^{(2,+)}(z)=E^{(2,+)}(z)\Phi^{\mathrm{(Ai)}}\left(|x|^{\frac{4}{3}}\eta(z)\right)
\left(s_0e^{2\pi i\alpha}\right)^{-\frac{\sigma_3}{2}}\sigma_{1}e^{\mp\pi i\alpha\sigma_3}z^{-\alpha\sigma_3}e^{-x^2g(z)\sigma_3},\quad \pm\Im z>0,
\end{equation}
where $\Phi^{\mathrm{(Ai)}}$ denotes the standard Airy parametrix (see Appendix \ref{AP} below), and $E^{(2,+)}(z)$ is given by
\begin{equation}\label{E2+}
E^{(2,+)}(z)=P^{(\infty)}(z)z^{\alpha\sigma_3}e^{\pm\pi i\alpha\sigma_3}\sigma_{1}\left(s_0e^{2\pi i\alpha}\right)^{\frac{\sigma_3}{2}}
\frac{1}{\sqrt{2}} \begin{pmatrix}1 & -i\\ -i &1\end{pmatrix}|x|^{\frac{\sigma_3}{3}}\eta(z)^{\frac{\sigma_3}{4}},\ \ \pm\Im z>0.
\end{equation}
Here, the branch of the function $\eta(z)^{\frac{1}{4}}$ is chosen such that $\arg\eta(z)\in(-\pi,\pi)$. This implies that on the segment $[z_{2,-},z_{2,+}]$, we have
\begin{equation}\label{etajump}
\left(\eta(z)^{\frac{\sigma_3}{4}}\right)_+=
\left(\eta(z)^{\frac{\sigma_3}{4}}\right)_-e^{\frac{\pi i}{2}\sigma_3}.
\end{equation}
It then follows from \eqref{Pinftyjump} and \eqref{etajump} that $E^{(2,+)}(z)$ is analytic in the neighborhood $U(z_{2,+},\delta)$. Finally, combining \eqref{Pinfty} and the asymptotic behavior \eqref{AiryAsyatinfty} with \eqref{P2+}, we get the matching condition \eqref{P2+matching}.

\subsection*{RH problem for $P^{(2,-)}(z)$}
\begin{description}
\item{(1)} $P^{(2,-)}(z)$ is analytic for $z\in U(z_{2,-},\delta)\setminus\Sigma_T$, where $U(z_{2,-},\delta)=\{z\in \mathbb{C}:|z-z_{2,-}|<\delta\}$.
\item{(2)} $P^{(2,-)}(z)$ shares the same jump conditions as $T(z)$ on $U(z_{2,-},\delta)\cap\Sigma_T$.
\item{(3)} On the boundary $\partial U(z_{2,-},\delta)=\{z\in \mathbb{C}:|z-z_{2,-}|=\delta\}$, we have
\begin{equation}\label{P2-matching}
P^{(2,-)}(z)=\left(I+O(|x|^{-2})\right)P^{(\infty)}(z),\quad \mathrm{as}\quad x\rightarrow -\infty.
\end{equation}
\end{description}
Similarly,
the solution to the above RH problem can  also be built out of the Airy function
\begin{equation}\label{P2-}
P^{(2,-)}(z)=E^{(2,-)}(z)\Phi^{(\mathrm{Ai})}\left(|x|^{\frac{4}{3}}\eta(-z)\right)
\left(s_0e^{2\pi i\alpha}\right)^{-\frac{\sigma_3}{2}}\sigma_2
z^{-\alpha\sigma_3}e^{-x^2g(z)\sigma_3},
\end{equation}
where $\Phi^{\mathrm{(Ai)}}$ again denotes the standard Airy  parametrix given in Appendix  \ref{AP} , $\eta(z)$ is defined in \eqref{eta} and $E^{(2,-)}(z)$ is given by
\begin{equation}\label{E2-}
E^{(2,-)}(z)=P^{(\infty)}(z)z^{\alpha\sigma_3}\sigma_2\left(s_0e^{2\pi i\alpha}\right)^{\frac{\sigma_3}{2}}
\frac{1}{\sqrt{2}} \begin{pmatrix}1 & -i\\ -i &1\end{pmatrix}|x|^{\frac{\sigma_3}{3}}\left\{\eta(-z)\right\}^{\frac{\sigma_3}{4}}.
\end{equation}
It is straightforward to check that $E^{(2,-)}(z)$ is analytic in the neighborhood $U(z_{2,-},\delta)$. The matching condition \eqref{P2-matching} follows from \eqref{Pinfty}, \eqref{AiryAsyatinfty} and \eqref{P2-}.

\subsection{Local parametrix near the origin }
In this subsection, we seek a parametrix  $P^{(0)}(z)$ satisfying the same jump conditions  as $\Sigma_T$ on the curves $\Sigma_{T}$ (see Figure \ref{Deforma3}) in the neighbourhood $U(0,\delta)$ of the origin and matching with $P^{(\infty)}(z)$ on the boundary $\partial U(0,\delta)$.

\subsection*{RH problem for $P^{(0)}(z)$}
\begin{description}
\item{(1)} $P^{(0)}(z)$ is analytic for $z\in U(0,\delta)\setminus\Sigma_T$, where $U(0,\delta)=\{z\in \mathbb{C}:|z|<\delta\}$.
\item{(2)} $P^{(0)}(z)$ satisfies the same jump conditions as $T(z)$ on $U(0,\delta)\cap\Sigma_T$.
\item{(3)} On the boundary $\partial U(0,\delta)=\{z\in \mathbb{C}:|z|=\delta\}$, we have
\begin{equation}\label{P0matching}
P^{(0)}(z)=\left(I+O(|x|^{-2})\right)P^{(\infty)}(z),\quad \mathrm{as}\quad x\rightarrow -\infty.
\end{equation}
\item{(4)} $P^{(0)}(z)$ has the same asymptotic behavior as $T(z)$ near the origin; see \eqref{T0}.
\end{description}
To proceed, we  define the conformal mapping
\begin{equation}\label{lambda}
\lambda(z)=\pm ig(z)=\pm\frac{1}{8}iz\left(z^2-\frac{8}{3}\right)^{\frac{3}{2}},\quad \pm\Im z>0,
\end{equation}
which has the following behavior at $z=0$
\begin{equation}
\lambda(z)=2^{\frac{3}{2}}3^{-\frac{3}{2}}z(1+o(1)),\quad \mathrm{as} \quad z\rightarrow0.
\end{equation}
Let $\Phi^{(\mathrm{Bes})}(z)$ be  the Bessel  paramatrix with parameter $\alpha$ as given in  Appendix \ref{Bessel}.  We then define
\begin{equation}\label{P0}
P^{(0)}(z)=E^{(0)}(z)\Phi^{(\mathrm{Bes})}\left(x^2\lambda(z)\right)C(z)
\left[s_1(e^{-2\pi i\alpha}+s_*)\right]^{-\frac{\sigma_3}{2}}
z^{-\alpha\sigma_3}e^{-x^2g(z)\sigma_3},
\end{equation}
where $\arg z\in(0,2\pi)$, $C(z)$ is a piecewise constant matrix defined in regions $\Lambda_k$ described in Figure \ref{Bes}
\begin{align*}
C(z)=\left\{\begin{aligned}
&I, \quad &z&\in \Lambda_1\cup\Lambda_8,\\
&\begin{pmatrix}1  & 0\\ -e^{-2\pi i\alpha} &1\end{pmatrix},\quad &z&\in \Lambda_2,\\
&\begin{pmatrix}e^{-\pi i\alpha}  & 0 \\ -e^{\pi i\alpha} & e^{\pi i\alpha}\end{pmatrix},\quad &z&\in \Lambda_3,\\
&e^{-\pi i\alpha\sigma_3},\quad &z&\in \Lambda_4,\\
&e^{\pi i\alpha\sigma_3},\quad &z&\in \Lambda_5,\\
&\begin{pmatrix}e^{\pi i\alpha}  & 0 \\ e^{-\pi i\alpha} & e^{-\pi i\alpha}\end{pmatrix},\quad &z&\in\Lambda_6,\\
&\begin{pmatrix}1  & 0\\ e^{2\pi i\alpha} & 1\end{pmatrix},\quad &z&\in \Lambda_7,
\end{aligned}\right.
\end{align*}
and $E^{(0)}(z)$ is given by
\begin{equation}\label{Ehat0}
E^{(0)}(z)=P^{(\infty)}(z)z^{\alpha\sigma_3}\left[s_1(e^{-2\pi i\alpha}+s_*)\right]^{\frac{\sigma_3}{2}}Q(z)e^{-\frac{1}{4}\pi i\sigma_3}\frac{1}{\sqrt{2}}\begin{pmatrix}1 & i\\ i & 1\end{pmatrix}
\end{equation}
with
\begin{equation}\label{Q}
Q(z)=\left\{\begin{aligned}
&e^{\frac{1}{2}\pi i\alpha\sigma_3}, \quad &\Im z>0,~~ |z|<\delta,\\
&\sigma_3\sigma_1 e^{\frac{1}{2}\pi i\alpha\sigma_3},\quad &\Im z<0,~~ |z|<\delta.
\end{aligned}\right.
\end{equation}

Using \eqref{Pinftyjump}, \eqref{Ehat0} and \eqref{Q},  we see that  $E^{(0)}(z)$ is analytic in the deleted neighborhood $U(0,\delta)\setminus \{0\}$.  Inserting  \eqref{Pinfty} into \eqref{Ehat0} shows that $E^{(0)}(z)$ is bounded at $z=0$. Therefore, $E^{(0)}(z)$ is analytic in the neighborhood $U(0,\delta)$.   Moreover, the matching condition \eqref{P0matching} follows from \eqref{P0}, \eqref{Ehat0} and \eqref{BesInfty}.

From \eqref{P0} and \eqref{Besseljump},   it is straightforward to verify that the function $P^{(0)}(z)$ constructed in \eqref{P0}  satisfies the same jump relations as $T(z)$ on $\Sigma_T\cap U(0,\delta) $.
Recalling the definition of the connection matrix in \eqref{E0}, we can rewrite the asymptotic behavior \eqref{T0} in the form
\begin{equation}\label{T1}
T(z)=\hat{T}_0(z)z^{\alpha\sigma_{3}}  \begin{pmatrix}1  & \frac{1}{1+e^{-2\pi i\alpha}}\\ 0&1\end{pmatrix} \begin{pmatrix}1  & 0\\ -e^{-2\pi i\alpha}&1\end{pmatrix}
\left[s_1(e^{-2\pi i\alpha}+s_*)\right]^{-\frac{\sigma_3}{2}}
z^{-\alpha\sigma_3}e^{-x^2g(z)\sigma_3},\end{equation}
where $\hat{T}_0(z)$ is analytic in a neighborhood of $z=0$.
Comparing \eqref{T1} with \eqref{P0}  and \eqref{BesParaExpand}, we see that $P^{(0)}(z)$  satisfies the asymptotic behavior \eqref{T0} as $z\to0$.

\subsection{Final transformation}
The final transformation is defined by
\begin{equation}\label{R}
R(z)=\left\{\begin{aligned}
&T(z)\left[P^{(1,\pm)}(z)\right]^{-1},\quad &z&\in U(z_{1,\pm},\delta)\setminus\Sigma_T,\\
&T(z)\left[P^{(2,\pm)}(z)\right]^{-1},\quad &z&\in U(z_{2,\pm},\delta)\setminus\Sigma_T,\\
&T(z)\left[P^{(0)}(z)\right]^{-1},\quad &z&\in U(0,\delta)\setminus\Sigma_T,\\
&T(z)\left[P^{(\infty)}(z)\right]^{-1},\quad  &\mathrm{e}&\mathrm{lsewhere}.\\ \end{aligned}
\right.
\end{equation}
Then, $R(z)$ satisfies the following RH problem.
\subsection*{RH problem for $R(z)$}

\begin{figure}[t]
  \centering
  \includegraphics[width=12cm,height=7cm]{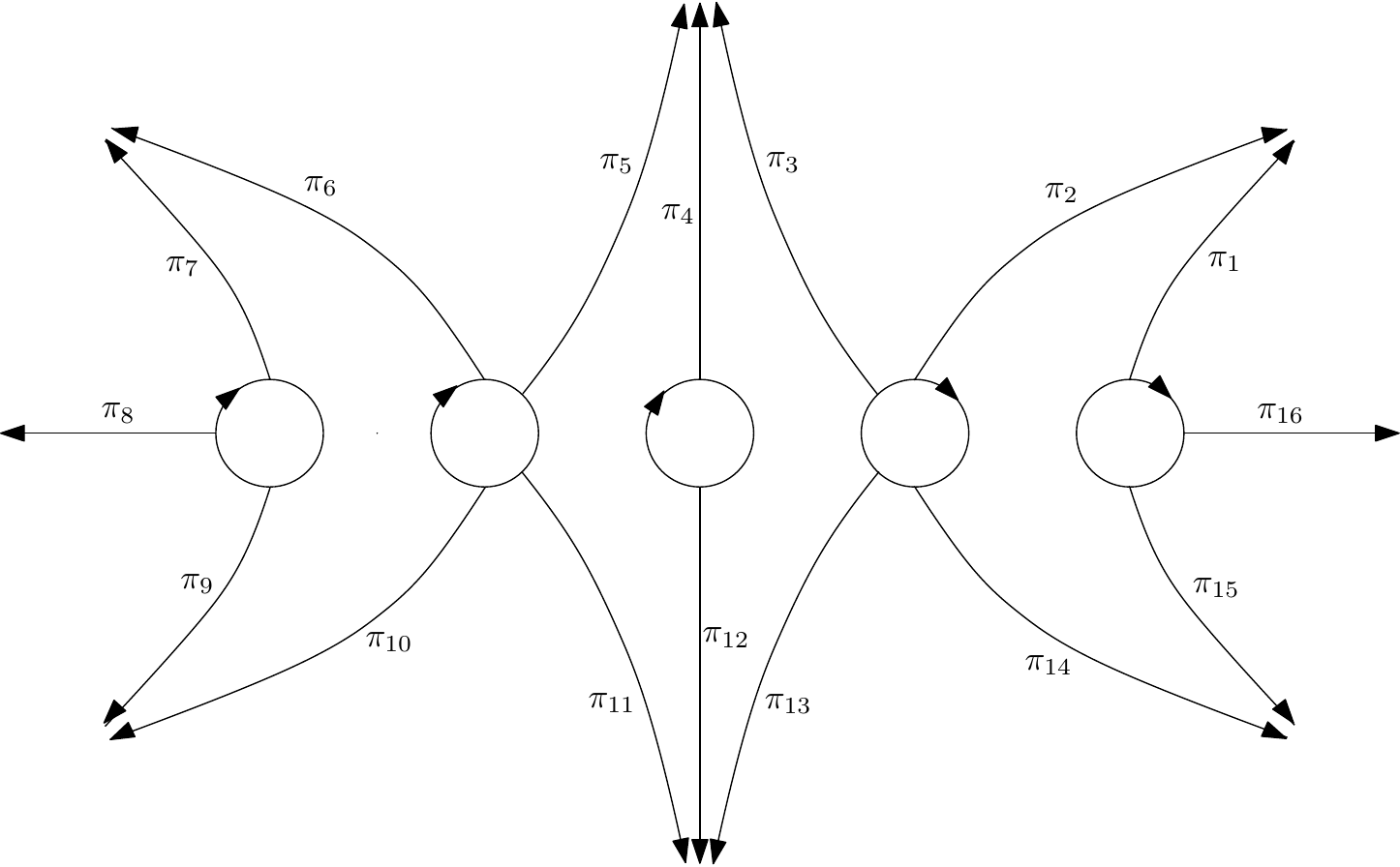}\\
  \caption{The jump contours $\Sigma_R$ of the RH problem for $R(z)$}\label{Rjump}
\end{figure}

\begin{description}
\item{(1)} $R(z)$ is analytic for $z\in \mathbb{C}\setminus\Sigma_R$, where the contour $\Sigma_R$ is illustrated in Figure \ref{Rjump}.
\item{(2)} On the contour $\Sigma_R$, we have $R_+(z)=R_-(z)J_{R}(z)$, where
 \begin{equation}\label{JumpR}
 J_{R}(z)=\left\{\begin{aligned}
&P^{(1,\pm)}(z)P^{(\infty)}(z)^{-1},\quad &z&\in \partial U(z_{1,\pm},\delta),\\
&P^{(2,\pm)}(z)P^{(\infty)}(z)^{-1},\quad &z&\in \partial U(z_{2,\pm},\delta),\\
&P^{(0)}(z)P^{(\infty)}(z)^{-1},\quad &z&\in \partial U(0,\delta),\\
&P^{(\infty)}(z)J_T(z)P^{(\infty)}(z)^{-1},\quad &z&\in \pi_k,\ k=1,\cdots,16.\\ \end{aligned}
\right.
 \end{equation}
\item{(3)} As $z\rightarrow\infty$, we have
\begin{equation}\label{Rexpan}
R(z)=I+\frac{R_1(x)}{z}+ O\left(\frac 1 {z^{2}}\right ).
\end{equation}
\end{description}
In view of the matching conditions \eqref{P1+matching}, \eqref{P1-matching}, \eqref{P2+matching}, \eqref{P2-matching}, and \eqref{P0matching}, it is readily seen that as $x\rightarrow -\infty$
\begin{equation}\label{JRestimation}
J_R(z)=\left\{\begin{aligned}
&I+O(|x|^{-2}), &z&\in\partial U(0,\delta)\cup\partial U(z_{1,\pm},\delta)\cup\partial U(z_{2,\pm},\delta),\\
&I+O(e^{-c_1|x|^2}),&z&\in\pi_k,\ k=1,\cdots,16,
\end{aligned}\right.
\end{equation}
where $c_1$ is a positive constant.

Consequently, we have
\begin{equation}\label{Restimation}
R(z)=I+O(|x|^{-2})\quad \mathrm{as} \quad x\rightarrow -\infty,
\end{equation}
 uniformly for $z\in \mathbb{C}\setminus\Sigma_R$.

\section{Proof of the main results }\label{sec:proof}

\subsection{Proof of Theorem \ref{thm}}\label{proof}
By \eqref{solu2}, the PIV solutions $q(x;\kappa )$ can be expressed in terms of
$\Phi_1$ in  \eqref{Asyatinfty1}.
Tracing back the series of transformations performed in Section \ref{RHanalysis}
$$\Phi\mapsto U\mapsto T\mapsto R,$$ we have that for large $z$
\begin{equation}\label{PhiRP}
e^{\frac{1}{3}x^2\sigma_3}\Phi(z)z^{-\alpha\sigma_3}e^{-x^2g(z)\sigma_3}
=R(z)P^{(\infty)}(z).
\end{equation}
We obtain the following expansion by using \eqref{Pinfty}, \eqref{H} and \eqref{repAB}
\begin{equation}\label{Pinftyatinfty}
P^{(\infty)}(z)=I+\frac{P^{(\infty)}_1(x)}{z}+O\left(\frac{1}{z^2}\right),\quad \mathrm{as} \quad z\rightarrow\infty,
\end{equation}
where
\begin{equation}\label{P1}
P_1^{(\infty)}(x)=\begin{pmatrix}
0 & -   \frac{ \sqrt{2}  ( c+ i\sqrt{2/3}\, ) }{ c+\sqrt{2}}
 s_0^{-1}D_{\infty}^{-2}\\
-   \frac{ \sqrt{2}  ( c+ i\sqrt{2/3}\, ) }{ c-\sqrt{2}}s_0D_{\infty}^{2} & 0
\end{pmatrix}.
\end{equation}
Substituting the expansions \eqref{Asyatinfty1}, \eqref{Rexpan} and \eqref{Pinftyatinfty} into \eqref{PhiRP}, we find
\begin{equation}\label{Phi1}
\Phi_1(x)=e^{-\frac 1 3 x^2 \sigma_3} \left (P_1^{(\infty)}(x)+R_1(x)\right ) e^{ \frac 1 3 x^2 \sigma_3} .
\end{equation}
In virtue of the error estimation \eqref{Restimation}, we  get
\begin{equation}\label{R1asymp}
R_1(x)=O(x^{-2}),\quad \mathrm{as} \quad x\rightarrow-\infty.
\end{equation}
Thus, we have
\begin{equation}\label{Phi1Asy}
\Phi_1(x)=e^{-\frac 1 3 x^2 \sigma_3} \left [P_1^{(\infty)}(x)+O\left(x^{-2}\right)\right ]e^{ \frac 1 3 x^2 \sigma_3} .
\end{equation}
Substituting the asymptotic approximation \eqref{Phi1Asy} into \eqref{solu2}, in view of
\eqref{P1}, we obtain
\begin{equation}\label{qexpre2}
q(x;\kappa )=-2x-4i\sqrt{\frac{2}{3}}\;\frac{c-i\sqrt{ {2}/{3}} }{c^2-2}x
+O(x^{-1}),\quad\mathrm{as}\quad x\rightarrow-\infty,
\end{equation}
where $c$ is defined in \eqref{c} and the error term is uniform for $x$ bounded away from the zeros of $c^2-2$.
Recalling the definition of $c$  in \eqref{c}, we may write
\begin{equation}\label{cexpre}
\frac{c-i\sqrt{ {2}/{3}}}{c^2-2}=i\frac{\sqrt{2}}{4\sqrt{3}}\frac{(2+e^{i\phi})(1+2e^{i\phi})}{1+e^{i\phi}+e^{2i\phi}}=i\frac{\sqrt{2}}{4\sqrt{3}}\left(2+\frac{3}{2\cos\phi+1}\right).
\end{equation}
By inserting \eqref{cexpre} and the expression  \eqref{c} of $\phi$  into   \eqref{qexpre2},  we arrive at the asymptotic expansion \eqref{qasymp-infty}. Finally, the connection formulas \eqref{connectionformula} follow from the relation   \eqref{Krepre} and the definition of $\nu_0$ in \eqref{nu}. This completes the proof of Theorem \ref{thm}.

\subsection{Proof of Corollary \ref{cor}}\label{proof:cor}
To derive the desired expansions for the poles,  we first recall the following
result for the zeros of real functions given in \cite{Hethcote}:
\begin{lem}\label{lem:zeros}
 In the interval $[z_h-\varrho, z_h+\varrho]$, suppose $f(x)=h(x)+\varepsilon(x)$,     where $f(x)$    is
continuous, $h(x)$ is differentiable, $h(z_h)=0$, $m=\min |h'(x)| >0$, and
   \begin{equation*}
   E_\varepsilon=\max |\varepsilon(x)|  < \min \{ |h(z_h-\varrho)|,  |h(z_h+\varrho)|\}.
 \end{equation*}
 Then there exists a zero $z_f$ of $f(x)$   in the interval such that
$|z_f-z_h|\leq E_\varepsilon/m$.\end{lem}

Now we are in a position to prove   Corollary \ref{cor}. First, use the same argument as in Section \ref{proof}, we have the asymptotic approximation
\begin{equation}\label{reciprocal-q}
\frac x{q(x; \kappa )} =\frac 3 4 \frac {2\cos\vartheta +1}{1-\cos\vartheta} +O\left (\frac {1}{ x^2}\right ),~~~x\to -\infty,
\end{equation}where $\vartheta=\vartheta(x)=\frac 1{\sqrt{3}} x^{2}-b\ln \left(2 \sqrt{3} x^{2}\right)+\psi$,  with $b$ and $\psi$  being given by \eqref{connectionformula}. The error term in \eqref{reciprocal-q} is uniform for $\vartheta$ bounded away from $2n\pi$ for integers $n$, as $x\to -\infty$. It is seen that $x\sim -3^{1/4} \vartheta^{1/2}$ and we may take $\vartheta$ as the large parameter. Applying Lemma \ref{lem:zeros}, we see that for large integers $n$, there exist zeros $x=a^{\pm}_n$ of $x/q(x;\kappa )$, corresponding respectively  to $\vartheta \sim 2n\pi \pm 2\pi/3$. More precisely, there exist poles $x=a^{\pm}_n$ of $q(x;\kappa )$ such that
\begin{equation*}
\vartheta(a^{\pm}_n)-\left(2n\pi \pm 2\pi/3\right )=O(1/ n),~~~~n\to\infty,
\end{equation*}from which \eqref{poles} follows.

\section*{Acknowledgements}
The work of Shuai-Xia Xu was supported in part by the National Natural Science Foundation of China under grant numbers 11571376 and 
 11971492, and by the Natural Science Foundation for Distinguished Young Scholars of Guangdong Province of China (Grant No.2022B1515020063).
 Yu-Qiu Zhao was supported in part by the National Natural Science Foundation of China under grant numbers 11571375 and 11971489.

\begin{appendices}
\section{Local parametrix models}
\subsection{Airy parametrix}\label{AP}
\begin{figure}[H]
  \centering
  \includegraphics[width=8cm,height=6cm]{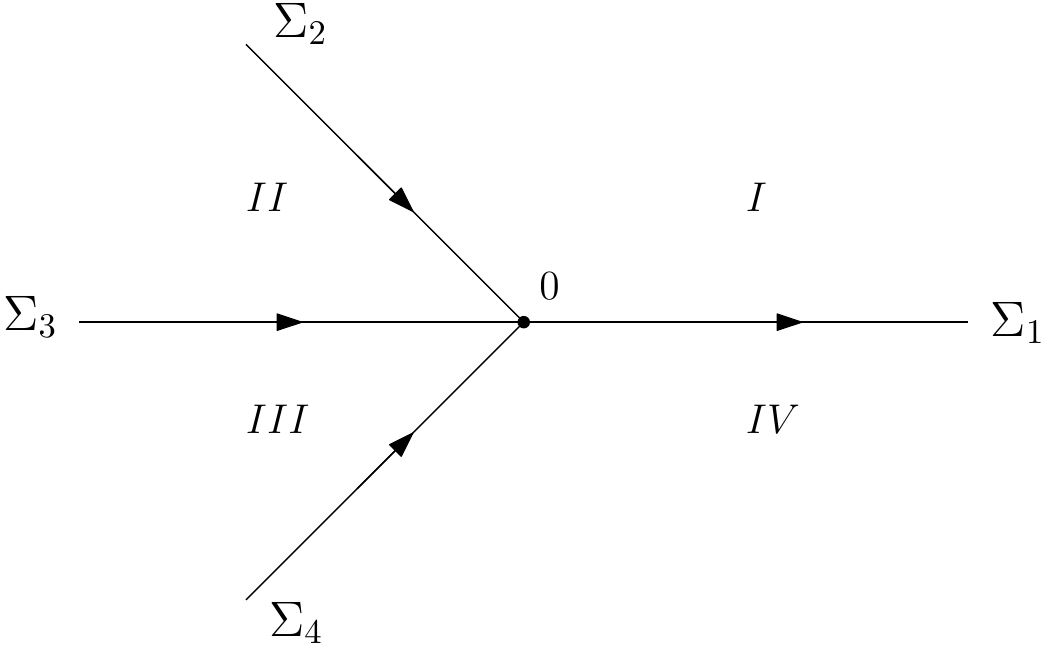}\\
  \caption{The jump contours and regions for $\Phi^{(\mathrm{Ai})}$}\label{Airy}
\end{figure}

Let $w=e^{2\pi i/3}$, we define
\begin{equation}
\Phi^{(\mathrm{Ai})}(s)=M\left\{
\begin{aligned}
&\begin{pmatrix}
\mathrm{Ai}(s) & \mathrm{Ai}(w^{2}s) \\
\mathrm{Ai}^{\prime}(s) & w^{2} \mathrm{Ai}^{\prime}(w^{2}s)
\end{pmatrix} e^{-i \frac{\pi}{6} \sigma_{3}}, &s&\in \mathrm{I},\\
&\begin{pmatrix}
\mathrm{Ai}(s) & \mathrm{Ai}(w^{2}s) \\
\mathrm{Ai}^{\prime}(s) & w^{2} \mathrm{Ai}^{\prime}(w^{2}s)
\end{pmatrix} e^{-i \frac{\pi}{6} \sigma_{3}}\begin{pmatrix}
1 & 0 \\
-1 & 1
\end{pmatrix},&s&\in \mathrm{II},\\
&\begin{pmatrix}
\mathrm{Ai}(s) & -w^{2}\mathrm{Ai}(ws) \\
\mathrm{Ai}^{\prime}(s) & -\mathrm{Ai}^{\prime}(w s)
\end{pmatrix} e^{-i \frac{\pi}{6} \sigma_{3}}\begin{pmatrix}
1 & 0 \\
1 & 1
\end{pmatrix},&s&\in \mathrm{III},\\
&\begin{pmatrix}
\mathrm{Ai}(s) & -w^{2} \mathrm{Ai}(w s) \\
\mathrm{Ai}^{\prime}(s) & -\mathrm{Ai}^{\prime}(ws)
\end{pmatrix} e^{-i \frac{\pi}{6} \sigma_{3}},&s&\in \mathrm{IV},
\end{aligned}
\right.
\end{equation}
where $\mathrm{Ai}(s)$ is the Airy function (cf. \cite[Chapter 9]{NIST}),
$$
M=\sqrt{2\pi}e^{\frac{1}{6}\pi i}\begin{pmatrix}1 & 0\\ 0 & -i \end{pmatrix},
$$
and the regions I-IV are shown in Figure \ref{Airy}. It is easy to check that $\Phi^{(\mathrm{Ai})}(s)$ solves the following RH problem (cf. \cite[Chapter 7]{Deft}):

\subsection*{RH problem for $\Phi^{(\mathrm{Ai})}(s)$}

\begin{description}
\item{(1)} $\Phi^\mathrm{(Ai)}(s)$ is analytic for $s\in \mathbb{C}\setminus \bigcup^4_{k=1}\Sigma_{k}$.
\item{(2)} $\Phi^\mathrm{(Ai)}(s)$ satisfies the jump relations $\Phi^\mathrm{(Ai)}_+(s)=\Phi^\mathrm{(Ai)}_-(s)J_k$, $s\in\Sigma_{k}$, $k=1,2,3,4$, where
\begin{equation*}
J_1=\begin{pmatrix}1 & 1 \\ 0 & 1 \end{pmatrix},\
J_2=\begin{pmatrix}1 & 0 \\ 1 & 1 \end{pmatrix}, \ J_3=\begin{pmatrix}0 & 1 \\ -1 & 0 \end{pmatrix}, \ J_4=\begin{pmatrix}1 & 0 \\ 1 & 1 \end{pmatrix}.
\end{equation*}
\item{(3)} $\Phi^\mathrm{(Ai)}(s)$ satisfies the following asymptotic behavior as $s\rightarrow \infty$:
\begin{equation}\label{AiryAsyatinfty}
\Phi^{(\mathrm{Ai})}(s)=s^{-\frac{\sigma_{3}}{4} }\frac{1}{\sqrt{2}} \begin{pmatrix}1 & i\\ i &1\end{pmatrix}\left(I+O\left(s^{-\frac{3}{2}}\right)\right) e^{-\frac{2}{3} s^{\frac{3}{2}} \sigma_{3}}.
\end{equation}
\end{description}

\subsection{Parabolic cylinder  parametrix}\label{PCP}

Let
$$\mathbf{ D}(s)=2^{-\frac{\sigma_{3}}{2}}\begin{pmatrix}
D_{-\nu-1}(i s) & D_{\nu}(s) \\[.2cm]
D_{-\nu-1}'(i s) & D_{\nu}'(s)\end{pmatrix}
\begin{pmatrix}
e^{i \frac{\pi}{2}(\nu+1)} & 0 \\
0 & 1
\end{pmatrix},
$$
where $D_{\nu}$ is the standard parabolic cylinder function with parameter $\nu$ (cf. \cite[Chapter 12]{NIST}).
Denote
$$
H_{0}=\begin{pmatrix}1 & 0 \\ h_{0} & 1\end{pmatrix}, \
H_{1}=\begin{pmatrix}1 & h_{1} \\ 0 & 1\end{pmatrix}, \
H_{n+2}=e^{i \pi\left(\nu+\frac{1}{2}\right) \sigma_{3}} H_{n} e^{-i \pi\left(\nu+\frac{1}{2}\right) \sigma_{3}}, \ n=0,1,
$$
with
\begin{equation}\label{h0}
h_{0}=-i \frac{\sqrt{2 \pi}}{\Gamma(\nu+1)}, \quad h_{1}=\frac{\sqrt{2 \pi}}{\Gamma(-\nu)} e^{i \pi \nu}, \quad 1+h_{0} h_{1}=e^{2 \pi i \nu}.
\end{equation}
We define
$$
\Phi^{(\mathrm{PC})}(s)=\left\{\begin{aligned}
&\mathbf{ D}(s),\quad && \arg s \in\left(- {\pi}/{4}, 0\right); \\
&\mathbf{ D}(s)H_0,\quad && \arg s \in\left(0,  {\pi}/{2}\right); \\
&\mathbf{ D}(s)H_1,\quad && \arg s \in\left( {\pi}/{2}, \pi\right); \\
&\mathbf{ D}(s)H_2,\quad && \arg s \in\left(\pi,  {3 \pi}/{2}\right); \\
&\mathbf{ D}(s)H_3,\quad && \arg s \in\left({3 \pi}/{2},  {7 \pi}/{4}\right).
\end{aligned}\right.
$$
Then $\Phi^{(\mathrm{PC})}(s)$ solves the following RH problem (cf. \cite{BI,FIKN}).

\subsection*{RH problem for $\Phi^{(\mathrm{PC})}(s)$}

\begin{figure}[H]
  \centering
  \includegraphics[width=7cm,height=6.1cm]{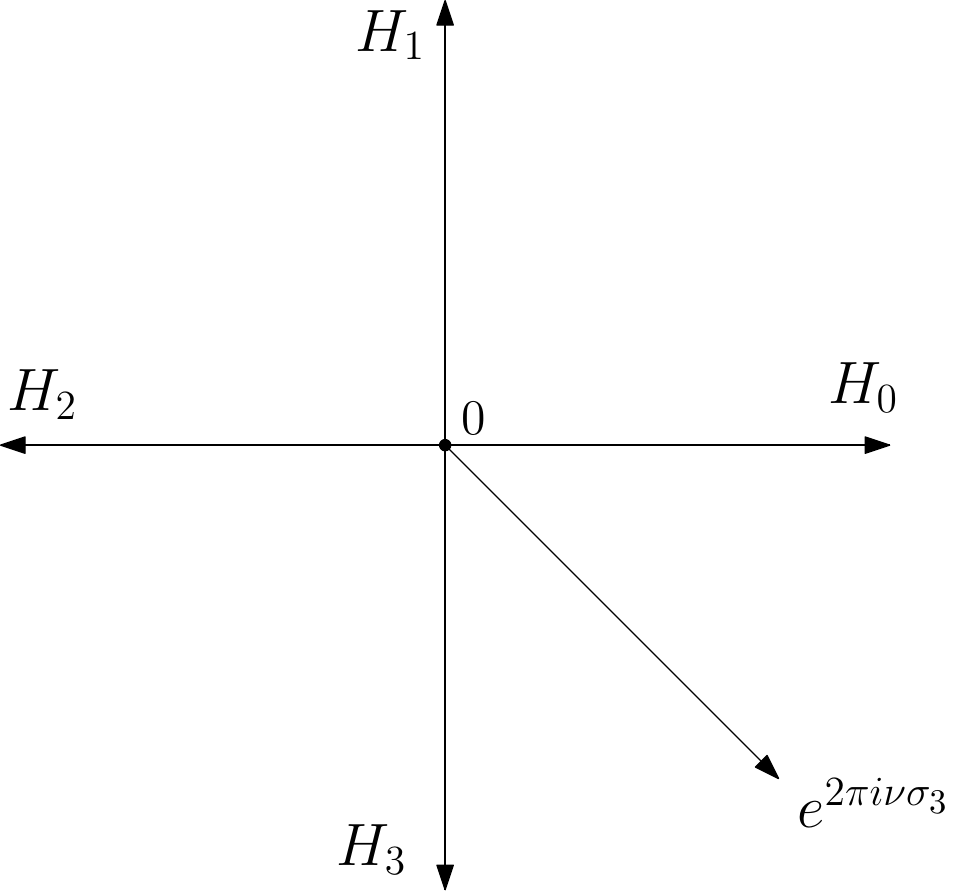}\\
  \caption{The jump contours and jump matrices for $\Phi^{(\mathrm{PC})}$}\label{PC}
\end{figure}

\begin{description}
\item{(1)} $\Phi^\mathrm{(PC)}(s)$ is analytic for all $s\in \mathbb{C}\setminus \bigcup^5_{k=1}\Sigma_{k}$, where $\Sigma_{k}=\{s\in\mathbb{C}:\arg s=\frac{k\pi}{2}\}$, $k=1,2,3,4$ and  $\Sigma_{5}=\{s\in\mathbb{C}:\arg s=-\frac{\pi}{4}\}$; see Figure \ref{PC}.
\item{(2)} $\Phi^\mathrm{(PC)}(s)$ satisfies the jump conditions as indicated in Figure \ref{PC}.
\item{(3)} $\Phi^\mathrm{(PC)}(s)$ has the following asymptotic behavior at infinity
\begin{align}\label{PCAsyatinfty}
\Phi^\mathrm{(PC)}(s)=\begin{pmatrix}0 &1 \\ 1 & -s\end{pmatrix}2^{\frac{\sigma_3}{2}}\begin{pmatrix}
1+O\left(\frac{1}{s^2}\right) & \frac{\nu}{s}+O\left(\frac{1}{s^3}\right) \\[.2cm] \frac{1}{s}+O\left(\frac{1}{s^3}\right) & 1+O\left(\frac{1}{s^2}\right)\end{pmatrix}
e^{\left(\frac{s^{2}}{4}-\nu\ln s\right) \sigma_{3}}.
\end{align}
\end{description}

\subsection{A Bessel model parametrix}\label{Bessel}

We start with the following RH problem.
\subsection*{RH problem for $\Phi^{(\mathrm{Bes})}(s)$}
\begin{figure}[H]
  \centering
  \includegraphics[width=6.8cm,height=6.8cm]{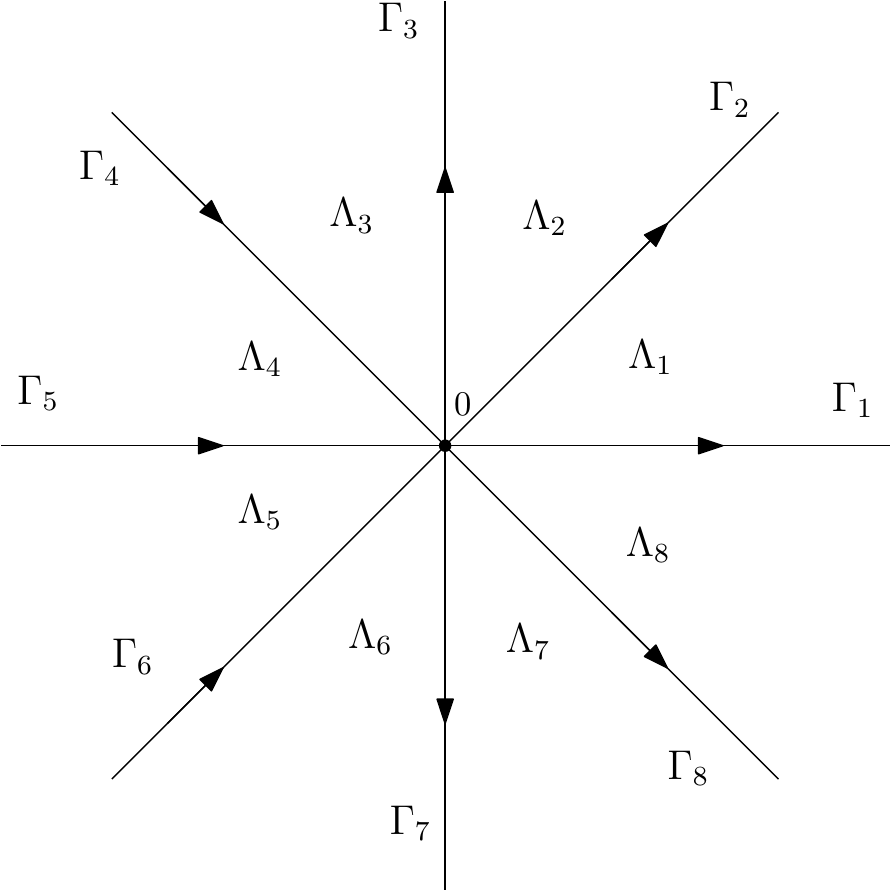}\\
  \caption{The jump contours and regions for $\Phi^{(\mathrm{Bes})}$}\label{Bes}
\end{figure}

\begin{description}
\item{(1)} $\Phi^\mathrm{(Bes)}(s)$ is analytic for all $s\in \mathbb{C}\setminus \bigcup^8_{k=1}\Gamma_{k}$, where $\Gamma_{k}=\{s\in \mathbb{C}:\arg s=k\pi/4\}$  are depicted in Figure \ref{Bes};
\item{(2)} $\Phi^\mathrm{(Bes)}(s)$ satisfies the following jump conditions
\begin{equation}\label{Besseljump}
\Phi_{+}^{(\mathrm{Bes})}(s)=\Phi_{-}^{(\mathrm{Bes})}(s)\left\{\begin{aligned} &\begin{pmatrix}0 & 1 \\ -1 & 0\end{pmatrix}, \quad &s&\in\Gamma_1\cup\Gamma_5,\\
&\begin{pmatrix}1 & 0 \\ e^{-2\pi i\alpha} & 1\end{pmatrix}, \quad &s&\in\Gamma_2\cup\Gamma_6,\\
&\begin{pmatrix}e^{\pi i\alpha} & 0 \\ 0 & e^{-\pi i\alpha}\end{pmatrix}, \quad  &s&\in\Gamma_3\cup\Gamma_7,\\
&\begin{pmatrix}1 & 0 \\ e^{2\pi i\alpha} & 1\end{pmatrix}, \quad  &s&\in\Gamma_4\cup\Gamma_8.
\end{aligned}\right.
\end{equation}

\item{(3)} The asymptotic behavior of $\Phi^\mathrm{(Bes)}(s)$ at infinity is different in each quadrant. As $s\rightarrow\infty$,
\begin{equation}\label{BesInfty}
\Phi^\mathrm{(Bes)}(s)=\frac{1}{\sqrt{2}}\begin{pmatrix}
1 & -i \\ -i & 1 \end{pmatrix}\left(I+O\left(s^{-1}\right)\right)e^{\frac{\pi i\sigma_3}{4}}e^{-is\sigma_3}\left\{
\begin{aligned}
&e^{-\frac{\alpha\pi i\sigma_3}{2}}, & s &\in \Lambda_1\cup\Lambda_2,\\
&e^{\frac{\alpha\pi i\sigma_3}{2}}, & s &\in \Lambda_3\cup\Lambda_4,\\
&e^{\frac{\alpha\pi i\sigma_3}{2}}\sigma_1\sigma_3, & s &\in \Lambda_5\cup\Lambda_6,\\
&e^{-\frac{\alpha\pi i\sigma_3}{2}}\sigma_1\sigma_3, & s &\in\Lambda_7\cup\Lambda_8.
\end{aligned}\right.
\end{equation}
\end{description}

According to \cite{Van}, the above RH problem can be constructed in terms of the modified Bessel function $I_{\alpha\pm\frac{1}{2}}(s)$ and $K_{\alpha\pm\frac{1}{2}}(s)$
\begin{equation}\label{BesPara}
\Phi^\mathrm{(Bes)}(s)=\begin{pmatrix}
\sqrt{\pi} s^{\frac{1}{2}}I_{\alpha+\frac{1}{2}}(se^{-\frac{\pi i}{2}}) & -\frac{1}{\sqrt{\pi}}s^{\frac{1}{2}}K_{\alpha+\frac{1}{2}}(se^{-\frac{\pi i}{2}})\\[.3cm]
-i\sqrt{\pi} s^{\frac{1}{2}}I_{\alpha-\frac{1}{2}}(se^{-\frac{\pi i}{2}})&-\frac{i}{\sqrt{\pi}}s^{\frac{1}{2}}K_{\alpha-\frac{1}{2}} (se^{-\frac{\pi i}{2}}  )
\end{pmatrix}e^{-\frac{1}{2}\alpha\pi i\sigma_3}\end{equation}
for  $s\in\Lambda_2$, where $s^{1/2}$ takes the principal branch.
While the explicit expressions of $\Phi^\mathrm{(Bes)}(z)$ in other sectors are determined by \eqref{BesPara}  and the  jump relation \eqref{Besseljump}.

Using the  series expansion of the  modified Bessel function  \cite[(10.25.2)]{NIST}
 $$I_{\nu}(s)=\left(\frac{1}{2}s\right )^{\nu}
 \sum_{k=0}^{\infty}\frac{\left(\frac{1}{4}s^2\right)^k}{k!\Gamma(\nu+k+1)}
 ~~~\mbox{for}~~\arg s\in (-\pi, \pi),$$
and the relation  \cite[(10.27.4)]{NIST}
$$K_{\nu}(s)=\frac{\pi}{2}\frac{I_{-\nu}(s)-I_{\nu}(s) }{\sin(\pi \nu)}, \quad \nu \not\in \mathbb{Z},$$
it is seen  from  \eqref{BesPara} that
\begin{equation}\label{BesParaExpand}
\Phi^\mathrm{(Bes)}(s)=B(s)\;s^{\alpha \sigma_3}\begin{pmatrix}
1 & \frac{1}{1+e^{-2\pi i \alpha}}\\
0&1
\end{pmatrix}, \quad s\in \Lambda_2,  \end{equation}
where $\alpha-\frac{1}{2} \not\in  \mathbb{Z} $ and  $B(s)$ is an entire function in  $s$.
 The behavior of $\Phi^\mathrm{(Bes)}(s)$ near the origin in the other sectors $\Lambda_k$ can be determined by \eqref{BesParaExpand} and the jump relations \eqref{Besseljump}.

\end{appendices}

\end{document}